\documentclass[floatfix,aps,twocolumn,showpacs,superscriptaddress,pra,10pt,longbibliography]{revtex4-2}
\usepackage{amsmath,amsfonts,amssymb,graphicx}
\usepackage{epstopdf}
\usepackage{bm}
\usepackage{hyperref}

\usepackage[utf8]{inputenc}
\usepackage[T1]{fontenc}
\usepackage[english]{babel}

\IfFileExists{newtxtext.sty}
    {\usepackage{newtxtext,newtxmath}}
    {\IfFileExists{stix.sty}
       {\usepackage{stix}}
       {\IfFileExists{mathptmx.sty}
       {\usepackage{mathptmx}}{} } }

\usepackage{hyperref}
\usepackage{xcolor}

\begin{document}
\title{Evolution of static and dynamical density correlations in a one-dimensional soft-core gas from the Tonks-Girardeau limit to a clustering fluid.}
\author{Martina Teruzzi}
\affiliation{Mathematics Area, mathLab, SISSA, International School for Advanced Studies, Via Bonomea 265, Trieste, Italy}
\affiliation{Dipartimento di Fisica ``Aldo Pontremoli'', Universit\`a degli Studi di Milano, via Celoria 16, I-20133 Milano, Italy}
\author{Christian Apostoli}
\affiliation{Dipartimento di Fisica ``Aldo Pontremoli'', Universit\`a degli Studi di Milano, via Celoria 16, I-20133 Milano, Italy}
\author{Davide Pini}
\affiliation{Dipartimento di Fisica ``Aldo Pontremoli'', Universit\`a degli Studi di Milano, via Celoria 16, I-20133 Milano, Italy}
\author{Davide Emilio Galli}
\affiliation{Dipartimento di Fisica ``Aldo Pontremoli'', Universit\`a degli Studi di Milano, via Celoria 16, I-20133 Milano, Italy}
\author{Gianluca Bertaina}
\affiliation{Istituto Nazionale di Ricerca Metrologica, Strada delle Cacce 91, I-10135 Torino, Italy}
\affiliation{Dipartimento di Fisica ``Aldo Pontremoli'', Universit\`a degli Studi di Milano, via Celoria 16, I-20133 Milano, Italy}
\date{\today}

\begin{abstract}
Repulsive soft-core atomic systems may undergo clustering if their density is high enough that core overlap is unavoidable. In one-dimensional quantum systems, it has been shown that this instability triggers a transition from a Luttinger liquid to various cluster Luttinger liquids. Here, we focus on the Luttinger liquid regime and theoretically study the evolution of key observables related to density fluctuations, that manifest a striking dependence on density. We tune the interaction so that the low-density regime corresponds to a Tonks-Girardeau gas, and show that as the density is increased the system departs more and more from Tonks-Girardeau behavior, displaying a much larger compressibility as well as rotonic excitations that finally drive the clustering transition. 
We compare various theoretical approaches, which are accurate in different regimes. Using quantum Monte Carlo methods and analytic continuation as a benchmark, we investigate the regime of validity of the mean-field Bogoliubov and the real-time multiconfiguration time-dependent Hartree-Fock approaches. Part of the behavior that we describe should be observable in ultracold Rydberg-dressed gases, provided that system losses are prevented. 
\end{abstract}

\maketitle

\section{Introduction}\label{sec:intro}

Particles interacting with repulsive soft-core potentials, which are flat at short distances, may manifest clustering when externally forced to relatively high densities, since the cores are bound to overlap and a potential energy gain can be obtained by forming groups of overlapping particles that repel each other. In the context of classical physics, a sufficient criterion was recognized for clustering, involving the Fourier transform of the potential \cite{likos_criterion_2001,mladek_formation_2006,Mladek_ComputerAssemblyClusterForming_2008,Lenz_MicroscopicallyResolvedSimulations_2012}. 

While, in classical mechanics, clustering favored by the potential competes with entropic effects at finite temperature, in the quantum regime one may expect an interesting phase diagram even at zero temperature, due to the role of zero-point motion. In fact, supersolid behavior, characterized by the coexistence of crystal and superfluid order, has been investigated for soft-core bosons \cite{henkel_threedimensional_2010,cinti_defectinduced_2014,saccani_excitation_2012,ancilotto_supersolid_2013,AngeloneSuperglassPhaseInteractionBlockaded2016,Prestipino_Freezingsoftcorebosons_2018}, and unconventional states have been predicted for soft-core fermions \cite{li_emergence_2016,Keles_wavesuperfluidityrepulsive_2020,Seydi_RydbergdressedFermiliquid_2021}. In one-dimensional (1D) systems, where the paradigm for quantum liquids is Luttinger liquid (LL) theory \cite{Giamarchi,Haldane}, tendency to clustering manifests as a transition to cluster Luttinger liquids (CLL), on a lattice \cite{mattioli_cluster_2013,dalmonte_cluster_2015} and in the continuum \cite{Rossotti_QuantumCriticalBehavior_2017}. 

In the context of ultracold gases, soft-core potentials are relevant when considering Rydberg-dressed gases~\cite{henkel_threedimensional_2010,pupillo_strongly_2010,balewski_rydberg_2014}, where single atoms are in a coherent superposition of their ground state and a highly excited Rydberg state~\cite{low_experimental_2012} obtained with off-resonant Rabi coupling. The resulting effective interaction between two Rydberg-dressed atoms is a soft shoulder potential, with a flat repulsive core of size $R_c$, related to the highly-excited orbital, and a repulsive van-der-Waals tail. Experiments are progressively increasing the coherence time of such systems \cite{Jau_Entanglingatomicspins_2016,zeiher_manybody_2016,Zeiher_CoherentManyBodySpin_2017}, although they are mostly focusing on fast out-of-equilibrium dynamics simulating Ising models~\cite{Borish_TransverseFieldIsingDynamics_2020,Guardado-Sanchez_QuenchDynamicsFermi_2021}, since equilibration is still a delicate issue due to decoherence to other Rydberg states. 
Although experimentally challenging, the regime in which soft-core potentials induce clustering is quite intriguing from the theoretical point of view. Clustering effects in zero temperature quantum systems are best investigated by considering the statistics of density fluctuations, which characterize the dynamical $S(q,\omega)$ and the static $S(q)$ structure factors. In particular, the peaks in $S(q,\omega)$ identify well-defined collective density fluctuation modes $\omega(q)$, which are phonons at small momenta and may display significant structure at higher momenta. In \cite{Rossotti_QuantumCriticalBehavior_2017}, the dynamical structure factor of a soft-core 1D bosonic system without dissipation was studied across the LL-to-CLL transition, driven by increasing interaction strength at a fixed specific density. The LL phase, far from being structureless, was found to manifest a roton excitation, in qualitative accordance with mean-field theory, similarly to the corresponding phases in higher dimensions \cite{henkel_threedimensional_2010,Macri_GroundStateExcitation_2014}. A secondary roton, accessible only with a fully ab initio quantum Monte Carlo (QMC) simulation, was argued to be the hallmark of an emergent quantum Ising model. For the same potential, in the weakly-interacting but high-density regime, Ref.~\cite{Prestipino_Clusterizationweaklyinteractingbosons_2019} showed that mean-field theory is quantitatively consistent with a variational approach based on a Gaussian expansion of the single-particle wavefunction around equally-spaced virtual cluster positions.

In the opposite regime of very low density, soft potentials are equivalent to the contact potential, and thus map to the Lieb-Liniger model~\cite{lieb_exact_1963a}. In particular, by suitably tuning the interaction strength, Ref.~\cite{QFS} showed that a Tonks-Girardeau (TG) regime \cite{Tonks,Girardeau} of impenetrable bosons, with its typical particle-hole flat dynamical structure factor, was accessible. In this article we explore the intriguing scenario of the evolution from a Tonks-Girardeau to a rotonic regime, and finally to a cluster phase, by only changing the density and keeping the interaction strength fixed. This is at variance with the pure Lieb-Liniger model in the hard-core limit, where the system remains in the TG regime whatever the density.

For the soft shoulder potential, we characterize this evolution by employing QMC methods complemented by analytic continuation of imaginary-time correlation functions, and analyze the limits of validity of approaches such as mean-field Bogoliubov theory, Feynman approximation, and the multiconfiguration time-dependent Hartree method.

In Sec.~\ref{sec:methods} we describe the Hamiltonian, the criterion for clustering and the theoretical methods that we employ to tackle the problem; in Sec.~\ref{sec:results}
we describe the results for the relevant observables, and in particular the spectrum of density fluctuations; in Sec.~\ref{sec:conclusions} we draw conclusions.

\section{Model and Methods}\label{sec:methods}
We consider a 1D system of $N$ bosons interacting with the following Hamiltonian in configuration space:
\begin{equation}\label{eq:hamiltonian}
 \mathcal{H} = -\frac{\hbar^2}{2m}\sum_i^N \Delta_{x_i} + \sum_{i<j} \frac{V_0}{r_{ij}^6+R_c^6} 
\end{equation}
where $m$ is the mass, $x_i$ is the coordinate of particle $i$ and $r_{ij}=|x_i-x_j|$ the distance of particles $i$ and $j$, $V_0$ and $R_c$ are the strength and the radius of the soft-core interaction potential. This is usually referred to as shoulder potential and has been considered as an effective model for Rydberg-dressed gases~\cite{henkel_threedimensional_2010,pupillo_strongly_2010,balewski_rydberg_2014} at density smaller than or of the order of $1/R_c$. We consider no hard-core part in the shoulder potential, assuming that realistic quasi-1D configurations effectively smooth it out, analogously to dilute gases modeled with the Lieb-Liniger Hamiltonian \cite{olshanii1998atomic,plodzien_rydberg_2017}. While we theoretically study also densities much higher than $1/R_c$, experimental applicability of our results in this regime is at the moment prevented due to increased losses, which can partially be reduced by using an underlying optical lattice. In the following we use $R_c$ as the unit for length, $1/R_c$ for wavevectors, $E_c=\hbar^2/m R_c^2$ for energy and $\hbar/E_c$ for time, leading to the dimensionless interaction strength $U=V_0/(R_c^6 E_c )$ and density $\rho=nR_c$. 

\begin{figure}[tbp]
\centering
\includegraphics[width = \columnwidth]{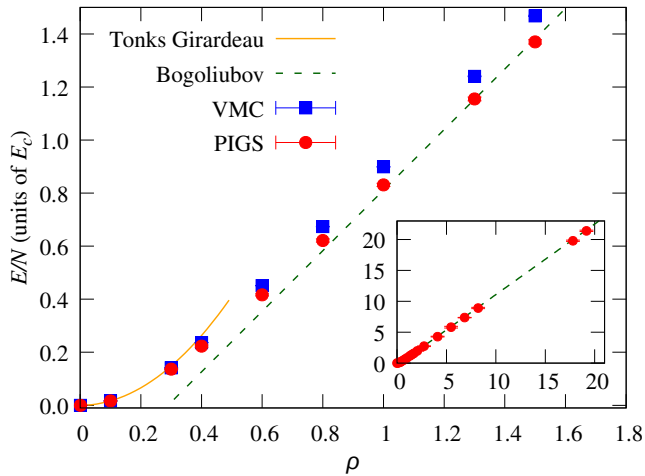}
\caption{Energy per particle as a function of the density from QMC simulations, compared with the Tonks--Girardeau model $E_{\text{TG}}/N$ (solid line) and to the shifted Bogoliubov result $E_B/N$ (dashed line). Inset: PIGS results in a wider range of densities.}
\label{fig:eq_state} 
\end{figure}

The Fourier transform of the potential \cite{StaticKorea} 
\begin{multline}
    \tilde{V}(q)\equiv\int V(x) e^{i q x}dx =\\
    \frac{U \pi}{3}  e^{-\frac{|q|}{2}} \left[\cos\left(\frac{|q|\sqrt{3}}{2}\right) + e^{-\frac{|q|}{2}} + \sqrt{3} \sin\left(\frac{|q|\sqrt{3}}{2}\right)\right]
\end{multline}
features a global minimum $\tilde{V}(q_c)<0$ at $q_c \simeq 4.3$, corresponding to a typical length $b_c=2\pi/q_c \simeq 1.46$. 
It has been shown that classically, even in the case of a completely repulsive potential, such a feature favors the formation of clusters at a mutual distance $\sim b_{c}$ \cite{likos_criterion_2001,mladek_formation_2006}. For the 1D system considered in the present case, an ordered phase is prevented by thermal fluctuations at all temperatures $T>0$ \cite{prestipino_cluster_2014,prestipino_probing_2015}, even though classical simulations observed and characterized distinct signatures of clustering tendency~\cite{Mambretti_EmergenceIsingcritical_2020}. When considering quantum effects at zero temperature, the cluster phase can be observed for large $U$ or high density, and it manifests itself as a cluster Luttinger liquid \cite{Rossotti_QuantumCriticalBehavior_2017,mattioli_cluster_2013,dalmonte_cluster_2015}.

\begin{figure}[btp]
\centering
\includegraphics[width = \columnwidth]{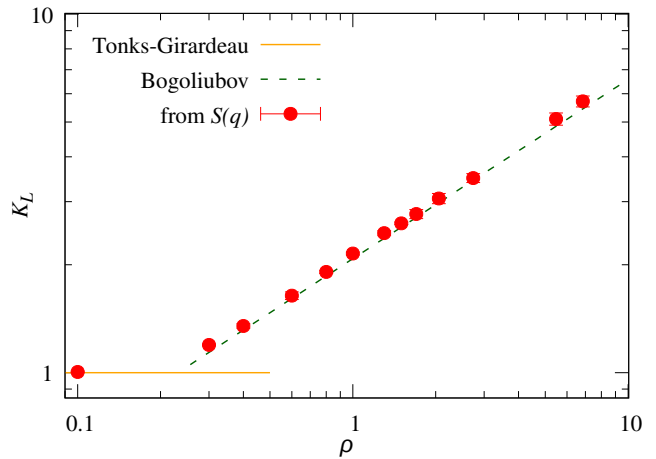}
\caption{Log-log plot of the Luttinger parameter $K_L$ as a function of the density $\rho$, extracted from the low-momentum behavior of the PIGS static structure factor, compared to the Tonks-Girardeau value $K_L=1$ and the Bogoliubov result (dashed line, see text).}
\label{fig:klut} 
\end{figure}

In the dilute limit $\rho\ll 1$, the macroscopic properties of the system depend only on the scattering length $a_{\text{1D}}$. This universality regime was verified in Ref.~\cite{QFS}, by showing that static and dynamical observables of the model \eqref{eq:hamiltonian} are compatible with those of the Lieb-Liniger~\cite{lieb_exact_1963a,caux_dynamical_2006} and hard-rods~\cite{motta_dynamical_2016} models, depending on whether the strength is smaller or, respectively, higher than $U=U_{{\text{TG}}}\simeq1.09$. This interaction corresponds to $a_{\text{1D}} \sim 0$, so that at very low density the system behaves as a TG gas.

Here we fix the interaction at $U=U_{{\text{TG}}}$ and increase the density from the Tonks-Girardeau gas to a relatively more compressible, but strongly correlated liquid, until we observe cluster formation. This behavior, different from the Lieb-Liniger model, stems from the shape of the potential, implying a role of both $n a_\text{1D}$ and $n R_c$. Using different theoretical approaches, we calculate the zero-temperature values of the energy per particle $E/N$, the pair distribution function $g_2(r)$, the dynamical $S(q,\omega) = \int dt e^{i \omega t} \langle e^{it\hat{H}}\hat{\rho}_q e^{-it\hat{H}} \hat{\rho}_{-q} \rangle/{(2\pi N)}$ and the static $S(q) = \langle \hat{\rho}_q \hat{\rho}_{-q} \rangle/N$ structure factors, where $\rho_q$ is the density operator in momentum space and $\langle\cdots\rangle$ stands for the ground-state expectation value.

\subsection{Bogoliubov approach}\label{subsec:Bogoliubov}
In the standard mean-field approach to a bosonic (quasi\babelhyphen{nobreak})condensate, the bosonic quantum field is replaced by a classical field $\Psi(x,t)$, and the Gross-Pitaevskii equation is obtained \cite{henkel_threedimensional_2010}:
\begin{equation}\label{eq:GP}
    i\frac{\partial\Psi(x,t)}{\partial t} = \left(-\frac{\Delta_x}{2} + \int V(x-x^\prime)\left|\Psi(x^\prime,t)\right|^2 dx^\prime \right) \Psi(x,t)
\end{equation}
By assuming a small perturbation with respect to a uniform condensate \begin{equation}
\Psi(x,t)=e^{-i\mu t}\left(\sqrt{\rho} +u e^{-i[ \varepsilon_B(q) t -q x]}-v^* e^{i [\varepsilon_B(q) t -q x]}\right),
\end{equation}
where $|u|$ and $|v|$ are much smaller than $\sqrt{\rho}$ and $\mu=\rho g_B \equiv\rho \tilde{V}(0)=2\pi \rho U/3$ is the ground-state chemical potential, one obtains the Bogoliubov de Gennes equations \cite{Macri_GroundStateExcitation_2014} that admit the analytical solution for the energy of excitations:
\begin{equation}\label{eq:bogo}
\varepsilon_B(q)=\sqrt{\varepsilon_0(q)\left[\varepsilon_0(q)+2 \rho \tilde{V}(q)\right]}  \; 
\end{equation}
where $\varepsilon_0(q)=q^2/2$ is the energy of a free particle. Notice that in this approach all properties depend on the combined coupling $\rho U$. However, given its assumption on the field, one expects it to be valid only for large enough densities \cite{Prestipino_Clusterizationweaklyinteractingbosons_2019}. In fact, the zero-density limit corresponds to the TG regime where surely the Bogoliubov approach breaks down. From Eq.~\eqref{eq:bogo} one may also infer the value of $\rho=\rho_{\text{CL}} \simeq 20.65/U$ at which the Bogoliubov dispersion becomes imaginary close to $q=q_c$, due to the dominance of the negative contribution from $\rho\tilde{V}(q)$, signalling the breakdown of this approximation and cluster formation~\cite{Rossotti_QuantumCriticalBehavior_2017,Prestipino_Clusterizationweaklyinteractingbosons_2019}. In our study, this corresponds to $\rho_{\text{CL}}\approx 18.9$.

\begin{figure}[tbp]
\centering
\includegraphics[width = \columnwidth]{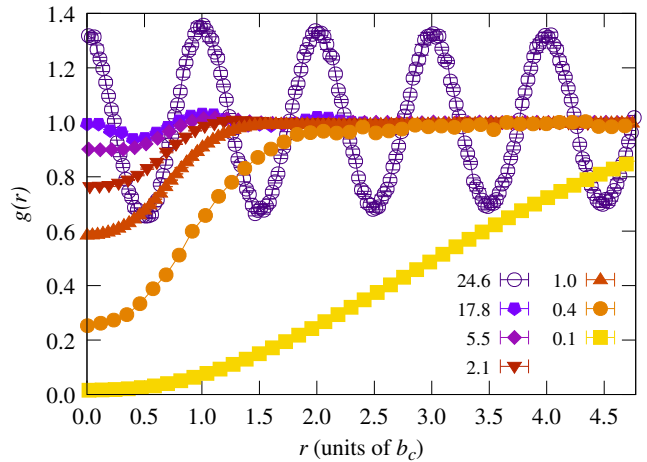}
\caption{Pair distribution function for increasing densities $\rho$ (symbols, PIGS). The distance between particles $r$ is in units of the typical clustering distance $b_c$, to highlight its role at high density. Lines are a guide to the eye.}
\label{fig:g2} 
\end{figure}

\subsection{Quantum Monte Carlo methods}\label{subsec:QMC}
We employ two canonical ensemble QMC methods: variational Monte Carlo (VMC) \cite{McMillan_GroundStateLiquid_1965} and Path Integral Ground State Monte Carlo (PIGS) \cite{PIGS,SPIGS}.
VMC evaluates expectation values $\langle\hat{\mathcal{O}}\rangle_{\text{VMC}}=\langle \Psi_{T}|\hat{\mathcal{O}}|\Psi_{T}\rangle/\langle \Psi_{T}|\Psi_{T}\rangle$ of various observables $\hat{\mathcal{O}}$ given an explicit trial wavefunction $\Psi_{T}$. The probability density to be sampled is $|\Psi_{T}(\{x_i\})|^2$, which depends on parameters to be optimized by minimizing the expectation value of energy or its variance. 
PIGS performs an imaginary-time evolution of the VMC trial state $|\Psi_\tau\rangle=\exp{(- \tau\hat{H})}|\Psi_{T}\rangle$. For sufficiently long projection, the ground state observables can be estimated with $\langle\hat{\mathcal{O}}\rangle_0 \simeq \langle \Psi_\tau|\hat{\mathcal{O}}|\Psi_\tau\rangle/\langle \Psi_\tau|\Psi_\tau\rangle$, provided the trial state has a non-negligible overlap with the ground state. We employ a fourth-order pair-Suzuki approximation of the density matrix and notice that, for bosons, this approach has been shown to be unbiased even if $\Psi_T(\{x_i\})=1$ was used \cite{Rossi_Exactgroundstate_2009}. In practice, using a trial wavefunction which has been previously optimized for VMC guarantees rapid convergence of the PIGS method.

We typically simulate $N=100$ particles (up to $N=360$ in the cluster phase) using periodic boundary conditions in a segment of length $L=N/\rho$. 
As in \cite{QFS,Rossotti_QuantumCriticalBehavior_2017}, we consider a two-body Jastrow form of the wavefunction: 
\begin{equation}
\Psi_T(\{x_i\}) = \exp\left\{-\frac{1}{2}\sum_{i<j} \left[u(r_{ij}) + \chi(r_{ij})\right]\right\}
\label{eq:wavefunction}
\end{equation}
where $u(r_{ij})$ and $\chi(r_{ij})$ are the short--range and the long--range contributions, respectively. This form would be exact if the  Hamiltonian could be diagonalized into the sum of two terms $H=H_{\text{rel}}+H_{\text{ph}}$, describing the short--range relative motion and the long--range phononic dynamics, respectively \cite{ReattoChester}.

\textit{Short--range}. The short--range correlation $\exp[-u(r)/2]$ is taken to be the solution of the two--body Schr\"{o}dinger equation with an auxiliary potential, with the boundary condition that the wave-function has zero derivative at a distance $\bar{R} < L/2$.
In the ultra--low density regime \cite{QFS}, we used a step auxiliary potential and tuned its height and radius so as to have the same 1D scattering length as for the shoulder potential. The resulting wavefunction is not suited to the moderate densities that we consider here: we therefore use a rescaled shoulder potential $V_{\text{eff}}(r)= s V(r/l)$, where the scaling parameters $s\simeq 1$ and $l\simeq 1$ are optimized with VMC. We also notice that for the stronger interactions that we considered in \cite{Rossotti_QuantumCriticalBehavior_2017}, a more refined auxiliary potential was needed, which included the mean-field effect of neighbouring clusters.

\textit{Long--range}. The long--range contribution to the Jastrow factor, $\chi(r)$ allows for the correct description of phonons and is taken to be of the Reatto--Chester form \cite{ReattoChester,QFS}: 
\begin{equation}
\chi(r) = - \frac{{\alpha}_r}{\beta} \log{\left[\frac{{\sinh}^{\beta}\left(\frac{\pi}{L k_c} \right) + {\sin}^{\beta} \left(\frac{\pi r}{L} \right)}{{\sinh}^{\beta}\left(\frac{\pi}{L k_c} \right) + 1}\right]} \; .
\label{longrange}
\end{equation}
where $k_c=2\pi/\bar{R}$ and we typically use $\beta=8$.

\begin{figure}[tbp]
\centering
\includegraphics[width = \columnwidth]{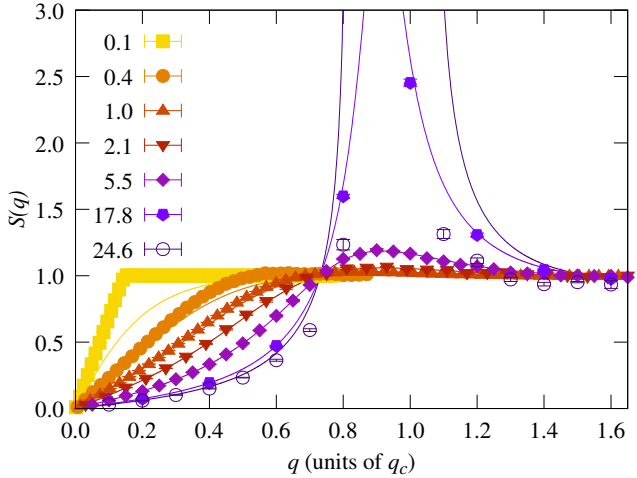}
\caption{Static structure factor for increasing densities $\rho$ (symbols, PIGS), compared to its Bogoliubov expression $S_B(q)$ (solid lines). Momentum is in units of $q_c$ to highlight its role at high density.}
\label{fig:staticSk} 
\end{figure}

Once a sufficiently long $\tau$ is considered, further projection allows to estimate the imaginary-time intermediate scattering function. We perform analytic continuation of this function with the Genetic Inversion via Falsification of Theories (GIFT) \cite{GIFT,bertaina_onedimensional_2016,Bertaina_Statisticalcomputationalintelligence_2017} algorithm, which gives access to the real-frequency dynamical structure factor. This algorithm is capable to reconstruct both narrow and broad spectral features \cite{motta_dynamical_2016}, so it is suited to investigate both the TG regime, dominated by particle-hole excitations, and the higher-density relatively more compressible regime, dominated by a single Bogoliubov excitation.

\subsection{Feynman approximation}\label{subsec:Feynman}

The zeroth momentum of the dynamical structure factor is $m_0(q)=\int_0^\infty S(q,\omega)d\omega= S(q)$. For a homogeneous system, the first momentum of the dynamical structure factor is determined by the $f$-sum rule $m_1(q)=\int_0^\infty \omega S(q,\omega)d\omega= \varepsilon_0(q)/\hbar$. These exact properties, together with the assumption that the dynamical structure factor is a delta function peaked at a single frequency, yields the Feynman approximation $S(q,\omega)=S(q)\delta(\omega-\varepsilon_{FA}(q)/\hbar)$  with $\varepsilon_{FA}(q)=\varepsilon_0(q)/S(q)$. We determine $\varepsilon_{FA}(q)$ using the static structure factor evaluated with the PIGS method described in the previous subsection.

The same single-mode approximation, together with the Bogoliubov dispersion of Eq.~\eqref{eq:bogo}, allows to derive a Bogoliubov expression for the static structure factor $S_B(q)=\varepsilon_{0}(q)/\varepsilon_{B}(q)$.

\begin{figure}[tbp]
\centering
\includegraphics[width = \columnwidth]{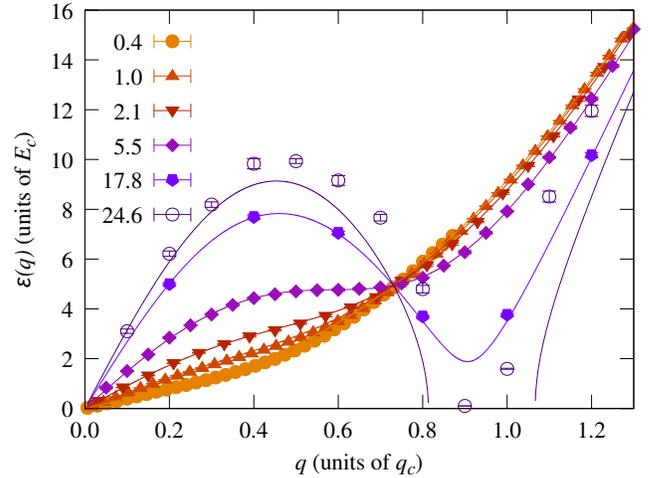}
\caption{Feynman approximation of the dispersion relation $\varepsilon_{FA}(q)$, using the PIGS results for the static structure factor, for increasing densities, compared to the Bogoliubov expression $\varepsilon_{B}(q)$ (solid lines). Momenta are in units of $q_c$ to highlight its role with increasing density.}
\label{fig:dispersion} 
\end{figure}

\subsection{Multiconfiguration time-dependent Hartree method}\label{subsec:MCTDH}
Real-time simulations are performed via the multiconfiguration time-dependent Hartree method (MCTDH) for identical particles~\cite{Meyer_MCTDH,Alon_MCTDH}, using the MCTDH-X software~\cite{MCTDH-X,Lin_2020,Lode_2016,Fasshauer_2016}. This technique approximates the single-particle Hilbert space by means of a basis of $M$ time-dependent orthonormal single-particle states $\left|\psi_k(t)\right\rangle$, which are called \textit{orbitals}. If we denote by $\left|n_1\dots n_M;t\right\rangle=\left|\mathbf{n},t\right\rangle$ the symmetrized product-state in which $n_k$ bosons occupy orbital $\left|\psi_k(t)\right\rangle$, then the MCTDH ansatz for a many-body state of $N$ particles is the following superposition:
\begin{equation}
\left|\phi (t)\right\rangle=\sum_\mathbf{n}C_\mathbf{n}(t)\left|\mathbf{n},t\right\rangle \; ,
\label{eq:MCTDH-ansatz}
\end{equation}
where the expansion coefficients $C_\mathbf{n}(t)$ are time-dependent, and the sum runs over all the symmetrized product-states of $N$ particles in $M$ orbitals. Plugging the ansatz~\eqref{eq:MCTDH-ansatz} into the time-dependent variational principle~\cite{Kramer,Kull} yields a coupled set of equations of motion for the coefficients $C_\mathbf{n}(t)$ and the orbitals $\left|\psi_k(t)\right\rangle$, whose solution gives the variationally optimized dynamics, which can be computed both in real and imaginary time. Note that the approximation lies in the dimensionality of the orbital basis, truncated to $M$, which is selected by physical considerations and computational resources' constraints \cite{master_thesis_Christian_Apostoli_2020}.

\section{Results}\label{sec:results}

In this section, we show the results of static properties such as the energy per particle, the Luttinger parameter, the pair distribution function and the static structure factor evaluated with QMC methods as a function of the density at fixed $U=U_{\text{TG}}$. Then, we approach the dynamics of density fluctuations using various techniques described in Sec.~\ref{sec:methods}: we evaluate the energy dispersion in the Bogoliubov and Feynman approaches, the dynamical structure factor from analytic continuation of imaginary-time PIGS simulations and, finally, we show the energy of low-lying excited states obtained by MCTDH simulations.

\begin{figure*}[tb]
\centering
\includegraphics[width = \textwidth]{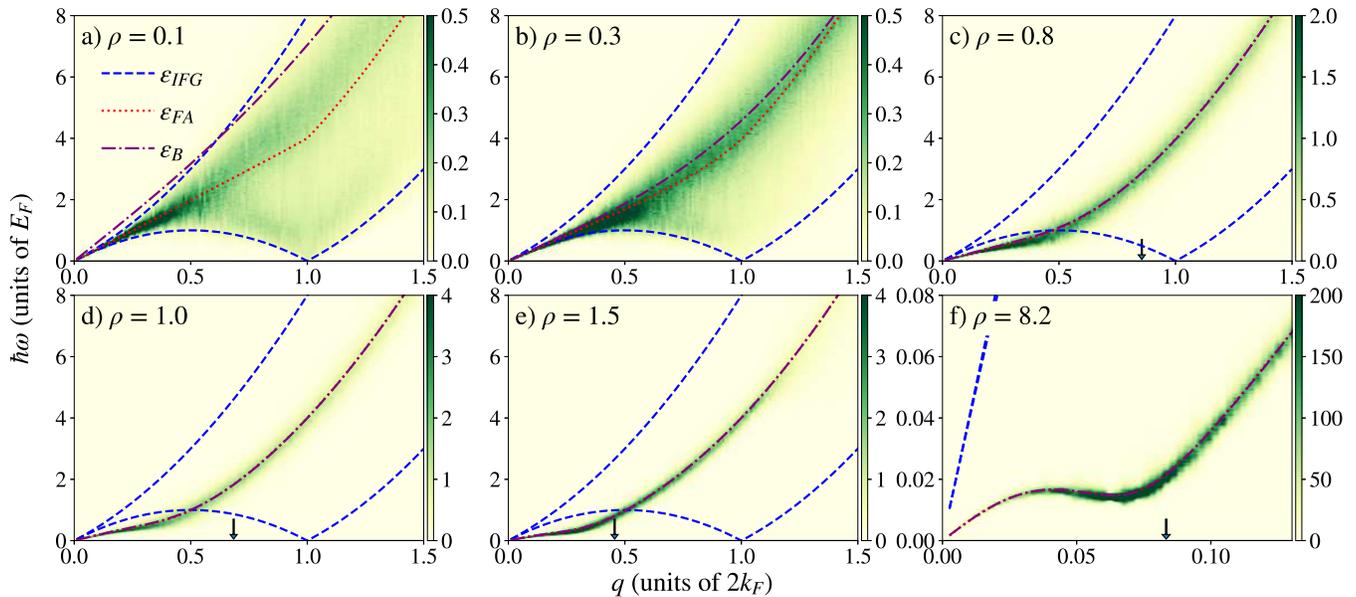}
\caption{(color online) Panels a-f: Dynamical structure factors for the densities $\rho=0.1, 0.3, 0.6, 0.8, 1.0, 1.5, 8.2$ from the analytic continuation of PIGS data, in units of the corresponding $\hbar/E_F$. The structure factors have been represented by a color map in the $(q,\omega)$ domain, with higher values of $S(q,\omega)$ corresponding to darker shades of green. The figure also shows the curves corresponding to the ideal Fermi gas particle-hole boundaries $\varepsilon_{IFG}$ (blue dashed lines) and to the Feynman sum-rule $\varepsilon_{FA}$ (red dotted line) and Bogoliubov $\varepsilon_{B}$ (purple dash-dotted line) approximations. In panels c-f, an arrow indicates the momentum $q=q_c$ in units of $2k_F$, and $\varepsilon_{B}$ essentially coincides with $\varepsilon_{FA}$. Notice that the cutoff of $S(q,\omega)$ is customized for each panel to highlight the dominant features and that panel f is magnified, to highlight the rotonic region.}
\label{fig:spectra} 
\end{figure*}

\subsection{Energy per particle and Luttinger parameter}\label{subsec:eos}

We compute the zero-temperature energy per particle for increasing density both with the VMC and the PIGS methods. In Fig.~\ref{fig:eq_state} we report the results compared with the TG result $E_{\text{TG}}/N=\pi^2\rho^2/6$. Only at very low density these two energies are compatible, while for $\rho\gtrsim 0.3$ they start to differ significantly. The increase in energy with density is slower than in the TG case, showing a relatively larger compressibility of the system. We ascribe this behavior to the fact that for increasing density the details of the shoulder potential start to matter and the absence of a hard core allows for particle overlap without large penalty. Indeed, the mean-field Bogoliubov equation of state $E_{B}/N=\rho g_B/2+e_0$, shifted by a constant $e_0\approx -0.33(1)E_c$ appears to be the main contribution to energy in the density range $1\lesssim \rho \lesssim 20$. More accurate expressions for the energy in the high-density regime have been investigated in \cite{Prestipino_Clusterizationweaklyinteractingbosons_2019}. 
Interestingly, the (non-shifted) Bogoliubov expression coincides with the mean-field, van der Waals approximation for the interaction contribution to the energy per particle of a {\em classical} 1D fluid with the same potential considered here, even though it should be recalled that at $T=0$ the classical system is actually a crystal at all densities and interaction strengths.

Let us now focus on the Luttinger parameter $K_L$ of 1D bosonization theory \cite{Giamarchi,Haldane}, which is related to the sound velocity $c$ of a homogeneous system via $c=\pi\rho/K_L$. The Luttinger parameter measures how compressible a fluid is with respect to an ideal Fermi gas at the same density, thus for the TG gas $K_L=1$, while $K_L>1$ indicates a more compressible fluid and $K_L<1$ a less compressible one such as the super Tonks-Girardeau or the hard-rod gases \cite{astrakharchik_tonks_2005,mazzanti_ground_2008,motta_dynamical_2016}. The value of $K_L$ can be derived from the energy per particle via the compressibility ${\kappa}^{-1} = \rho \frac{\partial}{\partial \rho}\left(\rho^2 \frac{\partial (E/N)}{\partial \rho}\right)$ and using the relation ${K_L}^2={\pi}^2{\rho}^3{\kappa}$. Moreover, it can be obtained from a linear fit of the static structure factor at small momenta using the relation $S(q)/q \underset{q\rightarrow0}{\rightarrow} K_L/(2\pi\rho)$ \cite{ReattoChester}. In particular, for the Reatto-Chester long-range Jastrow factor one finds the relation ${\alpha}_r=2/K_L$. 

In Fig.~\ref{fig:klut} we show the results from the fit of $S(k)$. For $\rho=0.1$ we are recovering the TG value $K_L=1$, as expected.
However, already at $\rho=0.3$ strong deviations from the TG model are visible. A weakly-interacting bosonic fluid behavior is confirmed by comparison to the Bogoliubov result $K_L=\pi\sqrt{\rho/g_{B}}=\sqrt{3\pi\rho/2U}$~\cite{Rossotti_QuantumCriticalBehavior_2017}.

\subsection{Pair distribution function}\label{subsec:gr}

In Figure~\ref{fig:g2} we show the evolution of the pair distribution function for increasing densities.
For $\rho=0.1$, $g(r)$ approaches the one expected for the ideal Fermi gas (IFG), as expected in the TG low-density limit. However, already at $\rho=0.3$ the pair distribution function is not zero anymore for $r\to0$. The small-distance depletion is gradually lost when the density increases, leading to a saturation in the large density limit and even to the manifestation of a local maximum at contact and then to significant oscillations in the cluster regime $\rho\gtrsim \rho_{\text{CL}}$. It is important to notice that such peaks appear at multiples $r=2\pi/q_c$, independent of density.

As already pointed out for the energy, also the behavior of $g(r)$ is similar to that brought out in former studies of its classical counterpart at $T\neq 0$: as $\rho$ is increased, at first $g(r)$ approaches $1$ over the whole $r$ range, meaning that correlations become weaker and weaker. This effect is due to the soft-core character of the interaction, which allows mutual overlap of the particles, thereby depressing density fluctuations, and is the reason why the mean-field approximation provides an accurate description of the fluid regime of the system at relatively high density. For soft-core interactions of the so-called $Q^{+}$ class~\cite{likos_criterion_2001}, i.e., featuring a monotonically decreasing Fourier transform, this loss of correlations would go on indefinitely, and the mean-field picture would become exact in the $\rho\rightarrow \infty$ limit~\cite{lang_fluid_2000}. Instead, for interactions belonging to the $Q^{\pm}$ class~\cite{likos_criterion_2001} as in the present case, whereby the Fourier transform presents an absolute minimum at nonzero wave vector, strong density modulations eventually appear as the stability limit of the fluid phase is approached, beyond which the cluster crystal takes over. 

\subsection{Static structure factor}\label{subsec:sq}

In Figure~\ref{fig:staticSk} we show the evolution of the static structure factor for increasing densities.
As for the pair distribution function, for $\rho=0.1$ the static structure factor approaches the IFG result: $S_{\text{IFG}}(q)=q/(2\pi\rho)$ for $q<2\pi\rho$, $S_{\text{IFG}}(q)=1$ for $q\ge 2\pi\rho$. At higher densities, the behavior of $S(q)$ departs more and more from the IFG result and is well described by the mean-field result $S_B(q)$, except for the simulation at the density $\rho=24.6$ in the cluster phase. The static structure factor shows higher peaks for increasing $\rho$, all located at $q=q_c$. In a finite system one cannot see a true divergence at $q=q_c$, but, in the harmonic approximation for the cluster phase~\cite{Rossotti_QuantumCriticalBehavior_2017}, the height of the peak scales with the number of bosons as $N^{1-2K_L/N_\text{CL}^2}$, where $N_\text{CL}$ is the number of particles per cluster.

\subsection{Dynamical Structure Factor}\label{subsec:sqw}

We discuss the collective density excitations of the system in three ways: we first compare the single-mode approximations of the Bogoliubov and  Feynman approaches, then we evaluate the full dynamical structure factor via analytic continuation of PIGS imaginary-time data, and finally we perform real-time simulations of long wavelength excitations via the MCTDH method.

For intermediate densities, far from the TG regime, but not so high to have clusters, we expect the Bogoliubov approximation to be valid. This hypothesis can be tested by comparing $\varepsilon_B(q)$ to the dispersion relation $\varepsilon_{FA}(q)$ which is obtained from the static structure factor calculated with the PIGS method. Fig.~\ref{fig:dispersion} shows that the two approaches are quantitatively consistent up to at least $\rho\simeq 18$, indicating that a single-mode approximation is indeed adequate.

Given the reliability of the Bogoliubov approximation for the dispersion, it is also easy to explain the apparent fixed point in the curves, once those are expressed in units of $q_c$ for momenta and $E_c$ for energy. It is clear that $\varepsilon_B(q)$ does not depend on density at momentum $q_0\simeq 3.14 q_c$, which corresponds to the first zero of $\tilde{V}(q)$. At $q=q_0$, then $\varepsilon_B(q_0)=\varepsilon_0(q_0)\simeq 4.94E_c$.

In Fig.~\ref{fig:spectra} we show the full dynamical structure factor as extracted from GIFT analytic continuation of imaginary-time PIGS correlation functions~\cite{Rossotti_QuantumCriticalBehavior_2017,Bertaina_Statisticalcomputationalintelligence_2017}.

In Ref.~\cite{QFS} we showed that $S(q,\omega)$ at $\rho=0.001$ and $U=U_\text{TG}$ was consistent, for each $q$, with a flat spectrum in between the two particle-hole boundaries $\varepsilon_{IFG}(q) = \left| k_F q \pm q^2/2\right|$, similarly to the TG model. In panel a of Fig.~\ref{fig:spectra}, we observe that at $\rho=0.1$, $S(q,\omega)$ has nonzero weight in the same region predicted by the TG model, but the energy excitation spectrum is more peaked around the Feynman approximation, showing the predominance of certain frequencies. Therefore, the dynamical structure factor is more sensitive than the pair distribution function and the static structure factor in signalling a departure from the low-density TG limit.

The accumulation of spectral weight close to a single mode is gradually more evident going to densities $\rho=0.3, 0.8$ (panels b,c). For higher densities (panels d-f) a single mode can be ascertained, with a residual width that is partially due to the analytic continuation method. In this Figure, differently form Fig.~\ref{fig:dispersion}, we scale frequency and momentum with the effective Fermi energy $E_F=\hbar^2\pi^2 n^2/2m$ and twice the Fermi momentum $2k_F=2\pi n$, respectively, which are the relevant scales in the TG regime. A small arrow in panels (c-f) marks the position of the special clustering momentum $q_c$. At density $\rho=8.2$ (panel f) a roton is apparent, consistently with the dispersion in Fig.~\ref{fig:dispersion}. We have not performed GIFT continuations at higher density, since high resolution at small momenta requires very large systems and long imaginary-time projections.

\subsection{Low-lying excited states via MCTDH}\label{subsec:real-time}

\begin{figure}[tb]
\centering
\includegraphics[width = \columnwidth]{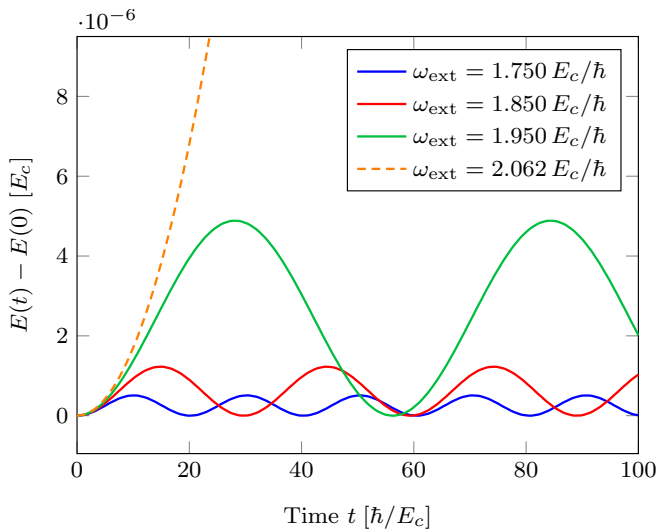}
\caption{Energy oscillations during the propagation of a 12-particle system at interaction $U=U_{\text{TG}}$ and density $\rho=2.05$, subject to the time-dependent perturbation~\eqref{eq:perturbation-pot} with amplitude $A=10^{-4}E_c$. The simulations with $\omega_{\textrm{ext}}= 1.750, 1.850, 1.950 \, E_c/\hbar$ show that, as $\omega_{\textrm{ext}}$ approaches $\omega(k_{\rm min})$, the oscillations grow in amplitude and decrease in frequency. If their frequencies $\omega_E$ are used to infer the excitation energy, the result for all three simulations is $\omega(k_{\rm min})\simeq2.062$. Indeed, a simulation with $\omega_{\textrm{ext}}=2.062$ (dashed line, only partially shown) exhibits no oscillation during the time of the simulation, but rather a steady energy growth.}
\label{fig:energy_osc} 
\end{figure}

We performed a series of MCTDH simulations of $N=12$ particles with up to $M=12$ orbitals in periodic boundary conditions. $N$ was limited by the computational costs; $M=12$ was deemed sufficient because the comparison of the MCTDH ground state energy with PIGS results shows very good agreement, and tests with higher $M$ yielded relative corrections to the energy of order $10^{-3}$ or lower, at the expense of a much higher computational cost. In the conditions studied here ($U=U_{\text{TG}}$), we observe that by increasing the density, the convergence with $M$ becomes faster. This is probably due to the system being relatively more compressible at higher density than in the TG regime, consistently with the adequacy of the Bogoliubov approximation.

To investigate the low-lying excited states, we simulate the real-time evolution of the system subject to a time-dependent perturbation potential of the form:
\begin{equation}
V_{\rm ext}(x,t) = A\sin(k_{\rm min}x-\omega_{\rm ext}t) \; ,
\label{eq:perturbation-pot}
\end{equation}
where the amplitude $A$ is much smaller than the interparticle interaction strength, $k_{\rm min} = 2\pi/L$ is the lowest wave vector compatible with periodic boundary conditions, $\omega_{\rm ext}$ is the probing frequency. As a consequence of the perturbation, the wavefunction of the system gains a small component on the excited state of momentum $k_{\rm min}$ and energy $E(k_{\rm min})$. If the perturbation was removed, the evolution of this superposition would conserve energy and would be periodic with frequency given by the energy difference between the excited state and the ground state: $\omega(k_{\rm min}) = (E(k_{\rm min})-E_{\rm gs})/\hbar$. However, in the persistence of the time-dependent perturbation, the energy is not conserved, but it increases when the oscillations of the system and the perturbation are in phase, and decreases when they are in phase opposition. Hence, the energy oscillates with frequency $\omega_E=|\omega_{\rm ext}-\omega(k_{\rm min})|$.

\begin{figure}[tb]
\centering
\includegraphics[width = \columnwidth]{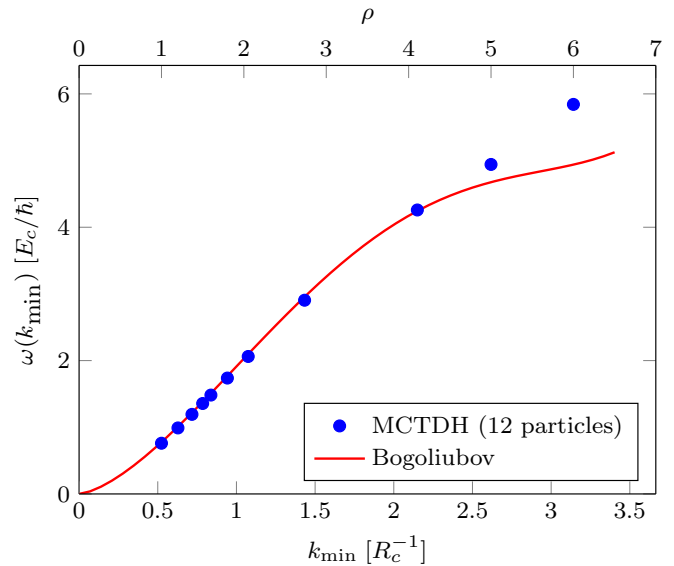}
\caption{Excitation energy of the low-lying excited states obtained by MCTDH simulations on a system of 12 particles, compared to the Bogoliubov spectrum. Note that these results have been obtained for systems with different densities, which are reported on the upper $x$-axis.}
\label{fig:lowlying} 
\end{figure}

The simulation procedure is as follows. The system is first relaxed over a static version ($\omega_{\rm ext}=0$) of the potential~\eqref{eq:perturbation-pot} to reduce initial transients. Then a real-time evolution with a probing frequency $\omega_{\rm ext}$ is performed. The frequency of the energy oscillations $\omega_E$ is measured, and the excitation energy is computed as $\omega(k_{\rm min})=\omega_{\rm ext}\pm\omega_E$. To determine the correct sign, a second simulation with a different $\omega_{\rm ext}$ is needed. We expect this method to work if the excitation spectrum exhibits a single phononic mode and if there are not non-phononic excitations at lower energy. Figure~\ref{fig:energy_osc} shows an example of this method which illustrates how the energy oscillations change as $\omega_{\rm ext}$ approaches $\omega(k_{\rm min})$.

Figure~\ref{fig:lowlying} shows the resulting $\omega(k_{\rm min})$. Please note that as $k_{\rm min}$ changes also the density $\rho$ changes, since $k_{\rm min} = 2\pi/L = 2\pi \rho/12$. For densities in the range to $1\le\rho\lesssim 3$, the Bogoliubov frequencies are in excellent agreement with the MCTDH results. Conversely, at higher densities the two curves become very different and the Bogoliubov values are lower than the MCTDH results. These differences are likely due to the size effects present in this small system of 12 particles, that become stronger for higher densities. 

\section{Conclusions}\label{sec:conclusions}
In summary, we  have shown that by varying the density, a 1D system of bosons with a suitably tuned shoulder interaction manifests a gradual crossover between a Tonks-Girardeau gas, where the density fluctuation spectrum is dominated by particle-hole excitations, to a relatively more compressible fluid dominated by a single Bogoliubov mode displaying a rotonic excitation close to a typical momentum independent of density. By using QMC results as a benchmark, we observed that mean-field theory is accurate in a wide range of densities, and devised a real-time protocol to study phononic excitations in this system with MCTDH. For even higher densities, a transition occurs towards large overlapping clusters. Future prospects include the characterization of this transition, which is expected to be different than the 1D quantum Ising universality class that is relevant to densities compatible with two-particle clusters~\cite{Rossotti_QuantumCriticalBehavior_2017}. Other interesting prospects are the study of trapped quasi-1D configurations~\cite{plodzien_rydberg_2017} with QMC and the investigation of quenches in real-time dynamics with MCTDH, possibly with fermions~\cite{Guardado-Sanchez_QuenchDynamicsFermi_2021}.

Data and scripts to reproduce the figures in this paper are available online (see Ref.~\cite{bertaina_zenodo_TG_2021}). 

\acknowledgments
We acknowledge the CINECA Awards IscraC-SOFTDYN (2015), IscraB-PANDA (2019) and IscraC-SEMIPRO (2019) for the availability of high performance computing resources and support. D.~P. acknowledges financial support by Universit\`a degli Studi di Milano, Project No. PSR2019\rule{0.15cm}{0.4pt}DIP\rule{0.15cm}{0.4pt}008-Linea~2. C. A. and D. E. G. acknowledge Axel U. J. Lode for the support in using the MCTDH-X software. 


\begin{thebibliography}{61}%
\makeatletter
\providecommand \@ifxundefined [1]{%
 \@ifx{#1\undefined}
}%
\providecommand \@ifnum [1]{%
 \ifnum #1\expandafter \@firstoftwo
 \else \expandafter \@secondoftwo
 \fi
}%
\providecommand \@ifx [1]{%
 \ifx #1\expandafter \@firstoftwo
 \else \expandafter \@secondoftwo
 \fi
}%
\providecommand \natexlab [1]{#1}%
\providecommand \enquote  [1]{``#1''}%
\providecommand \bibnamefont  [1]{#1}%
\providecommand \bibfnamefont [1]{#1}%
\providecommand \citenamefont [1]{#1}%
\providecommand \href@noop [0]{\@secondoftwo}%
\providecommand \href [0]{\begingroup \@sanitize@url \@href}%
\providecommand \@href[1]{\@@startlink{#1}\@@href}%
\providecommand \@@href[1]{\endgroup#1\@@endlink}%
\providecommand \@sanitize@url [0]{\catcode `\\12\catcode `\$12\catcode
  `\&12\catcode `\#12\catcode `\^12\catcode `\_12\catcode `\%12\relax}%
\providecommand \@@startlink[1]{}%
\providecommand \@@endlink[0]{}%
\providecommand \url  [0]{\begingroup\@sanitize@url \@url }%
\providecommand \@url [1]{\endgroup\@href {#1}{\urlprefix }}%
\providecommand \urlprefix  [0]{URL }%
\providecommand \Eprint [0]{\href }%
\providecommand \doibase [0]{https://doi.org/}%
\providecommand \selectlanguage [0]{\@gobble}%
\providecommand \bibinfo  [0]{\@secondoftwo}%
\providecommand \bibfield  [0]{\@secondoftwo}%
\providecommand \translation [1]{[#1]}%
\providecommand \BibitemOpen [0]{}%
\providecommand \bibitemStop [0]{}%
\providecommand \bibitemNoStop [0]{.\EOS\space}%
\providecommand \EOS [0]{\spacefactor3000\relax}%
\providecommand \BibitemShut  [1]{\csname bibitem#1\endcsname}%
\let\auto@bib@innerbib\@empty
\bibitem [{\citenamefont {Likos}\ \emph {et~al.}(2001)\citenamefont {Likos},
  \citenamefont {Lang}, \citenamefont {Watzlawek},\ and\ \citenamefont
  {L{\"o}wen}}]{likos_criterion_2001}%
  \BibitemOpen
  \bibfield  {author} {\bibinfo {author} {\bibfnamefont {C.~N.}\ \bibnamefont
  {Likos}}, \bibinfo {author} {\bibfnamefont {A.}~\bibnamefont {Lang}},
  \bibinfo {author} {\bibfnamefont {M.}~\bibnamefont {Watzlawek}},\ and\
  \bibinfo {author} {\bibfnamefont {H.}~\bibnamefont {L{\"o}wen}},\ }\bibfield
  {title} {\bibinfo {title} {Criterion for determining clustering versus
  reentrant melting behavior for bounded interaction potentials},\ }\href
  {https://doi.org/10.1103/PhysRevE.63.031206} {\bibfield  {journal} {\bibinfo
  {journal} {Phys. Rev. E}\ }\textbf {\bibinfo {volume} {63}},\ \bibinfo
  {pages} {031206} (\bibinfo {year} {2001})}\BibitemShut {NoStop}%
\bibitem [{\citenamefont {Mladek}\ \emph {et~al.}(2006)\citenamefont {Mladek},
  \citenamefont {Gottwald}, \citenamefont {Kahl}, \citenamefont {Neumann},\
  and\ \citenamefont {Likos}}]{mladek_formation_2006}%
  \BibitemOpen
  \bibfield  {author} {\bibinfo {author} {\bibfnamefont {B.~M.}\ \bibnamefont
  {Mladek}}, \bibinfo {author} {\bibfnamefont {D.}~\bibnamefont {Gottwald}},
  \bibinfo {author} {\bibfnamefont {G.}~\bibnamefont {Kahl}}, \bibinfo {author}
  {\bibfnamefont {M.}~\bibnamefont {Neumann}},\ and\ \bibinfo {author}
  {\bibfnamefont {C.~N.}\ \bibnamefont {Likos}},\ }\bibfield  {title} {\bibinfo
  {title} {Formation of {{Polymorphic Cluster Phases}} for a {{Class}} of
  {{Models}} of {{Purely Repulsive Soft Spheres}}},\ }\href
  {https://doi.org/10.1103/PhysRevLett.96.045701} {\bibfield  {journal}
  {\bibinfo  {journal} {Phys. Rev. Lett.}\ }\textbf {\bibinfo {volume} {96}},\
  \bibinfo {pages} {045701} (\bibinfo {year} {2006})}\BibitemShut {NoStop}%
\bibitem [{\citenamefont {Mladek}\ \emph {et~al.}(2008)\citenamefont {Mladek},
  \citenamefont {Kahl},\ and\ \citenamefont
  {Likos}}]{Mladek_ComputerAssemblyClusterForming_2008}%
  \BibitemOpen
  \bibfield  {author} {\bibinfo {author} {\bibfnamefont {B.~M.}\ \bibnamefont
  {Mladek}}, \bibinfo {author} {\bibfnamefont {G.}~\bibnamefont {Kahl}},\ and\
  \bibinfo {author} {\bibfnamefont {C.~N.}\ \bibnamefont {Likos}},\ }\bibfield
  {title} {\bibinfo {title} {Computer {{Assembly}} of {{Cluster}}-{{Forming
  Amphiphilic Dendrimers}}},\ }\href
  {https://doi.org/10.1103/PhysRevLett.100.028301} {\bibfield  {journal}
  {\bibinfo  {journal} {Phys. Rev. Lett.}\ }\textbf {\bibinfo {volume} {100}},\
  \bibinfo {pages} {028301} (\bibinfo {year} {2008})}\BibitemShut {NoStop}%
\bibitem [{\citenamefont {Lenz}\ \emph {et~al.}(2012)\citenamefont {Lenz},
  \citenamefont {Blaak}, \citenamefont {Likos},\ and\ \citenamefont
  {Mladek}}]{Lenz_MicroscopicallyResolvedSimulations_2012}%
  \BibitemOpen
  \bibfield  {author} {\bibinfo {author} {\bibfnamefont {D.~A.}\ \bibnamefont
  {Lenz}}, \bibinfo {author} {\bibfnamefont {R.}~\bibnamefont {Blaak}},
  \bibinfo {author} {\bibfnamefont {C.~N.}\ \bibnamefont {Likos}},\ and\
  \bibinfo {author} {\bibfnamefont {B.~M.}\ \bibnamefont {Mladek}},\ }\bibfield
   {title} {\bibinfo {title} {Microscopically {{Resolved Simulations Prove}}
  the {{Existence}} of {{Soft Cluster Crystals}}},\ }\href
  {https://doi.org/10.1103/PhysRevLett.109.228301} {\bibfield  {journal}
  {\bibinfo  {journal} {Phys. Rev. Lett.}\ }\textbf {\bibinfo {volume} {109}},\
  \bibinfo {pages} {228301} (\bibinfo {year} {2012})}\BibitemShut {NoStop}%
\bibitem [{\citenamefont {Henkel}\ \emph {et~al.}(2010)\citenamefont {Henkel},
  \citenamefont {Nath},\ and\ \citenamefont
  {Pohl}}]{henkel_threedimensional_2010}%
  \BibitemOpen
  \bibfield  {author} {\bibinfo {author} {\bibfnamefont {N.}~\bibnamefont
  {Henkel}}, \bibinfo {author} {\bibfnamefont {R.}~\bibnamefont {Nath}},\ and\
  \bibinfo {author} {\bibfnamefont {T.}~\bibnamefont {Pohl}},\ }\bibfield
  {title} {\bibinfo {title} {Three-{{Dimensional Roton Excitations}} and
  {{Supersolid Formation}} in {{Rydberg}}-{{Excited Bose}}-{{Einstein
  Condensates}}},\ }\href {https://doi.org/10.1103/PhysRevLett.104.195302}
  {\bibfield  {journal} {\bibinfo  {journal} {Phys. Rev. Lett.}\ }\textbf
  {\bibinfo {volume} {104}},\ \bibinfo {pages} {195302} (\bibinfo {year}
  {2010})}\BibitemShut {NoStop}%
\bibitem [{\citenamefont {Cinti}\ \emph {et~al.}(2014)\citenamefont {Cinti},
  \citenamefont {Macr{\`i}}, \citenamefont {Lechner}, \citenamefont {Pupillo},\
  and\ \citenamefont {Pohl}}]{cinti_defectinduced_2014}%
  \BibitemOpen
  \bibfield  {author} {\bibinfo {author} {\bibfnamefont {F.}~\bibnamefont
  {Cinti}}, \bibinfo {author} {\bibfnamefont {T.}~\bibnamefont {Macr{\`i}}},
  \bibinfo {author} {\bibfnamefont {W.}~\bibnamefont {Lechner}}, \bibinfo
  {author} {\bibfnamefont {G.}~\bibnamefont {Pupillo}},\ and\ \bibinfo {author}
  {\bibfnamefont {T.}~\bibnamefont {Pohl}},\ }\bibfield  {title} {\bibinfo
  {title} {Defect-induced supersolidity with soft-core bosons},\ }\href
  {https://doi.org/10.1038/ncomms4235} {\bibfield  {journal} {\bibinfo
  {journal} {Nat. Commun.}\ }\textbf {\bibinfo {volume} {5}},\ \bibinfo {pages}
  {3235} (\bibinfo {year} {2014})}\BibitemShut {NoStop}%
\bibitem [{\citenamefont {Saccani}\ \emph {et~al.}(2012)\citenamefont
  {Saccani}, \citenamefont {Moroni},\ and\ \citenamefont
  {Boninsegni}}]{saccani_excitation_2012}%
  \BibitemOpen
  \bibfield  {author} {\bibinfo {author} {\bibfnamefont {S.}~\bibnamefont
  {Saccani}}, \bibinfo {author} {\bibfnamefont {S.}~\bibnamefont {Moroni}},\
  and\ \bibinfo {author} {\bibfnamefont {M.}~\bibnamefont {Boninsegni}},\
  }\bibfield  {title} {\bibinfo {title} {Excitation {{Spectrum}} of a
  {{Supersolid}}},\ }\href {https://doi.org/10.1103/PhysRevLett.108.175301}
  {\bibfield  {journal} {\bibinfo  {journal} {Phys. Rev. Lett.}\ }\textbf
  {\bibinfo {volume} {108}},\ \bibinfo {pages} {175301} (\bibinfo {year}
  {2012})}\BibitemShut {NoStop}%
\bibitem [{\citenamefont {Ancilotto}\ \emph {et~al.}(2013)\citenamefont
  {Ancilotto}, \citenamefont {Rossi},\ and\ \citenamefont
  {Toigo}}]{ancilotto_supersolid_2013}%
  \BibitemOpen
  \bibfield  {author} {\bibinfo {author} {\bibfnamefont {F.}~\bibnamefont
  {Ancilotto}}, \bibinfo {author} {\bibfnamefont {M.}~\bibnamefont {Rossi}},\
  and\ \bibinfo {author} {\bibfnamefont {F.}~\bibnamefont {Toigo}},\ }\bibfield
   {title} {\bibinfo {title} {Supersolid structure and excitation spectrum of
  soft-core bosons in three dimensions},\ }\href
  {https://doi.org/10.1103/PhysRevA.88.033618} {\bibfield  {journal} {\bibinfo
  {journal} {Phys. Rev. A}\ }\textbf {\bibinfo {volume} {88}},\ \bibinfo
  {pages} {033618} (\bibinfo {year} {2013})}\BibitemShut {NoStop}%
\bibitem [{\citenamefont {Angelone}\ \emph {et~al.}(2016)\citenamefont
  {Angelone}, \citenamefont {Mezzacapo},\ and\ \citenamefont
  {Pupillo}}]{AngeloneSuperglassPhaseInteractionBlockaded2016}%
  \BibitemOpen
  \bibfield  {author} {\bibinfo {author} {\bibfnamefont {A.}~\bibnamefont
  {Angelone}}, \bibinfo {author} {\bibfnamefont {F.}~\bibnamefont
  {Mezzacapo}},\ and\ \bibinfo {author} {\bibfnamefont {G.}~\bibnamefont
  {Pupillo}},\ }\bibfield  {title} {\bibinfo {title} {Superglass {{Phase}} of
  {{Interaction}}-{{Blockaded Gases}} on a {{Triangular Lattice}}},\ }\href
  {https://doi.org/10.1103/PhysRevLett.116.135303} {\bibfield  {journal}
  {\bibinfo  {journal} {Phys. Rev. Lett.}\ }\textbf {\bibinfo {volume} {116}},\
  \bibinfo {pages} {135303} (\bibinfo {year} {2016})}\BibitemShut {NoStop}%
\bibitem [{\citenamefont {Prestipino}\ \emph {et~al.}(2018)\citenamefont
  {Prestipino}, \citenamefont {Sergi},\ and\ \citenamefont
  {Bruno}}]{Prestipino_Freezingsoftcorebosons_2018}%
  \BibitemOpen
  \bibfield  {author} {\bibinfo {author} {\bibfnamefont {S.}~\bibnamefont
  {Prestipino}}, \bibinfo {author} {\bibfnamefont {A.}~\bibnamefont {Sergi}},\
  and\ \bibinfo {author} {\bibfnamefont {E.}~\bibnamefont {Bruno}},\ }\bibfield
   {title} {\bibinfo {title} {Freezing of soft-core bosons at zero temperature:
  {{A}} variational theory},\ }\href
  {https://doi.org/10.1103/PhysRevB.98.104104} {\bibfield  {journal} {\bibinfo
  {journal} {Phys. Revi. B}\ }\textbf {\bibinfo {volume} {98}},\ \bibinfo
  {pages} {104104} (\bibinfo {year} {2018})}\BibitemShut {NoStop}%
\bibitem [{\citenamefont {Li}\ \emph {et~al.}(2016)\citenamefont {Li},
  \citenamefont {Hsieh}, \citenamefont {Mou},\ and\ \citenamefont
  {Wang}}]{li_emergence_2016}%
  \BibitemOpen
  \bibfield  {author} {\bibinfo {author} {\bibfnamefont {W.-H.}\ \bibnamefont
  {Li}}, \bibinfo {author} {\bibfnamefont {T.-C.}\ \bibnamefont {Hsieh}},
  \bibinfo {author} {\bibfnamefont {C.-Y.}\ \bibnamefont {Mou}},\ and\ \bibinfo
  {author} {\bibfnamefont {D.-W.}\ \bibnamefont {Wang}},\ }\bibfield  {title}
  {\bibinfo {title} {Emergence of a {{Metallic Quantum Solid Phase}} in a
  {{Rydberg}}-{{Dressed Fermi Gas}}},\ }\href
  {https://doi.org/10.1103/PhysRevLett.117.035301} {\bibfield  {journal}
  {\bibinfo  {journal} {Phys. Rev. Lett.}\ }\textbf {\bibinfo {volume} {117}},\
  \bibinfo {pages} {035301} (\bibinfo {year} {2016})}\BibitemShut {NoStop}%
\bibitem [{\citenamefont {Kele{\c s}}\ \emph {et~al.}(2020)\citenamefont
  {Kele{\c s}}, \citenamefont {Zhao},\ and\ \citenamefont
  {Li}}]{Keles_wavesuperfluidityrepulsive_2020}%
  \BibitemOpen
  \bibfield  {author} {\bibinfo {author} {\bibfnamefont {A.}~\bibnamefont
  {Kele{\c s}}}, \bibinfo {author} {\bibfnamefont {E.}~\bibnamefont {Zhao}},\
  and\ \bibinfo {author} {\bibfnamefont {X.}~\bibnamefont {Li}},\ }\bibfield
  {title} {\bibinfo {title} {$f$-wave superfluidity from repulsive interaction
  in {{Rydberg}}-dressed {{Fermi}} gas},\ }\href
  {https://doi.org/10.1103/PhysRevA.101.023624} {\bibfield  {journal} {\bibinfo
   {journal} {Phys. Rev. A}\ }\textbf {\bibinfo {volume} {101}},\ \bibinfo
  {pages} {023624} (\bibinfo {year} {2020})}\BibitemShut {NoStop}%
\bibitem [{\citenamefont {Seydi}\ \emph {et~al.}(2021)\citenamefont {Seydi},
  \citenamefont {Abedinpour}, \citenamefont {Asgari}, \citenamefont
  {Panholzer},\ and\ \citenamefont
  {Tanatar}}]{Seydi_RydbergdressedFermiliquid_2021}%
  \BibitemOpen
  \bibfield  {author} {\bibinfo {author} {\bibfnamefont {I.}~\bibnamefont
  {Seydi}}, \bibinfo {author} {\bibfnamefont {S.~H.}\ \bibnamefont
  {Abedinpour}}, \bibinfo {author} {\bibfnamefont {R.}~\bibnamefont {Asgari}},
  \bibinfo {author} {\bibfnamefont {M.}~\bibnamefont {Panholzer}},\ and\
  \bibinfo {author} {\bibfnamefont {B.}~\bibnamefont {Tanatar}},\ }\bibfield
  {title} {\bibinfo {title} {Rydberg-dressed {{Fermi}} liquid: {{Correlations}}
  and signatures of droplet crystallization},\ }\href
  {https://doi.org/10.1103/PhysRevA.103.043308} {\bibfield  {journal} {\bibinfo
   {journal} {Phys. Rev. A}\ }\textbf {\bibinfo {volume} {103}},\ \bibinfo
  {pages} {043308} (\bibinfo {year} {2021})}\BibitemShut {NoStop}%
\bibitem [{\citenamefont {Giamarchi}(2003)}]{Giamarchi}%
  \BibitemOpen
  \bibfield  {author} {\bibinfo {author} {\bibfnamefont {T.}~\bibnamefont
  {Giamarchi}},\ }\href
  {https://doi.org/10.1093/acprof:oso/9780198525004.001.0001} {\emph {\bibinfo
  {title} {Quantum Physics in One Dimension}}}\ (\bibinfo  {publisher}
  {{Oxford} {University} {Press}},\ \bibinfo {year} {2003})\BibitemShut
  {NoStop}%
\bibitem [{\citenamefont {Haldane}(1981)}]{Haldane}%
  \BibitemOpen
  \bibfield  {author} {\bibinfo {author} {\bibfnamefont {F.~D.~M.}\
  \bibnamefont {Haldane}},\ }\bibfield  {title} {\bibinfo {title} {Effective
  harmonic--fluid approach to low--energy properties of one--dimensional
  quantum fluids},\ }\href {https://doi.org/10.1103/PhysRevLett.47.1840}
  {\bibfield  {journal} {\bibinfo  {journal} {{Phys}. {Rev}. {Lett}.}\ }\textbf
  {\bibinfo {volume} {47}},\ \bibinfo {pages} {1840} (\bibinfo {year}
  {1981})}\BibitemShut {NoStop}%
\bibitem [{\citenamefont {Mattioli}\ \emph {et~al.}(2013)\citenamefont
  {Mattioli}, \citenamefont {Dalmonte}, \citenamefont {Lechner},\ and\
  \citenamefont {Pupillo}}]{mattioli_cluster_2013}%
  \BibitemOpen
  \bibfield  {author} {\bibinfo {author} {\bibfnamefont {M.}~\bibnamefont
  {Mattioli}}, \bibinfo {author} {\bibfnamefont {M.}~\bibnamefont {Dalmonte}},
  \bibinfo {author} {\bibfnamefont {W.}~\bibnamefont {Lechner}},\ and\ \bibinfo
  {author} {\bibfnamefont {G.}~\bibnamefont {Pupillo}},\ }\bibfield  {title}
  {\bibinfo {title} {Cluster {{Luttinger Liquids}} of {{Rydberg}}-{{Dressed
  Atoms}} in {{Optical Lattices}}},\ }\href
  {https://doi.org/10.1103/PhysRevLett.111.165302} {\bibfield  {journal}
  {\bibinfo  {journal} {Phys. Rev. Lett.}\ }\textbf {\bibinfo {volume} {111}},\
  \bibinfo {pages} {165302} (\bibinfo {year} {2013})}\BibitemShut {NoStop}%
\bibitem [{\citenamefont {Dalmonte}\ \emph {et~al.}(2015)\citenamefont
  {Dalmonte}, \citenamefont {Lechner}, \citenamefont {Cai}, \citenamefont
  {Mattioli}, \citenamefont {Läuchli},\ and\ \citenamefont
  {Pupillo}}]{dalmonte_cluster_2015}%
  \BibitemOpen
  \bibfield  {author} {\bibinfo {author} {\bibfnamefont {M.}~\bibnamefont
  {Dalmonte}}, \bibinfo {author} {\bibfnamefont {W.}~\bibnamefont {Lechner}},
  \bibinfo {author} {\bibfnamefont {Z.}~\bibnamefont {Cai}}, \bibinfo {author}
  {\bibfnamefont {M.}~\bibnamefont {Mattioli}}, \bibinfo {author}
  {\bibfnamefont {A.~M.}\ \bibnamefont {Läuchli}},\ and\ \bibinfo {author}
  {\bibfnamefont {G.}~\bibnamefont {Pupillo}},\ }\bibfield  {title} {\bibinfo
  {title} {Cluster {{Luttinger}} liquids and emergent supersymmetric conformal
  critical points in the one-dimensional soft-shoulder {{Hubbard}} model},\
  }\href {https://doi.org/10.1103/PhysRevB.92.045106} {\bibfield  {journal}
  {\bibinfo  {journal} {Phys. Rev. B}\ }\textbf {\bibinfo {volume} {92}},\
  \bibinfo {pages} {045106} (\bibinfo {year} {2015})}\BibitemShut {NoStop}%
\bibitem [{\citenamefont {Rossotti}\ \emph {et~al.}(2017)\citenamefont
  {Rossotti}, \citenamefont {Teruzzi}, \citenamefont {Pini}, \citenamefont
  {Galli},\ and\ \citenamefont
  {Bertaina}}]{Rossotti_QuantumCriticalBehavior_2017}%
  \BibitemOpen
  \bibfield  {author} {\bibinfo {author} {\bibfnamefont {S.}~\bibnamefont
  {Rossotti}}, \bibinfo {author} {\bibfnamefont {M.}~\bibnamefont {Teruzzi}},
  \bibinfo {author} {\bibfnamefont {D.}~\bibnamefont {Pini}}, \bibinfo {author}
  {\bibfnamefont {D.~E.}\ \bibnamefont {Galli}},\ and\ \bibinfo {author}
  {\bibfnamefont {G.}~\bibnamefont {Bertaina}},\ }\bibfield  {title} {\bibinfo
  {title} {Quantum {{Critical Behavior}} of {{One}}-{{Dimensional Soft Bosons}}
  in the {{Continuum}}},\ }\href
  {https://doi.org/10.1103/PhysRevLett.119.215301} {\bibfield  {journal}
  {\bibinfo  {journal} {Phys. Rev. Lett.}\ }\textbf {\bibinfo {volume} {119}},\
  \bibinfo {pages} {215301} (\bibinfo {year} {2017})}\BibitemShut {NoStop}%
\bibitem [{\citenamefont {Pupillo}\ \emph {et~al.}(2010)\citenamefont
  {Pupillo}, \citenamefont {Micheli}, \citenamefont {Boninsegni}, \citenamefont
  {Lesanovsky},\ and\ \citenamefont {Zoller}}]{pupillo_strongly_2010}%
  \BibitemOpen
  \bibfield  {author} {\bibinfo {author} {\bibfnamefont {G.}~\bibnamefont
  {Pupillo}}, \bibinfo {author} {\bibfnamefont {A.}~\bibnamefont {Micheli}},
  \bibinfo {author} {\bibfnamefont {M.}~\bibnamefont {Boninsegni}}, \bibinfo
  {author} {\bibfnamefont {I.}~\bibnamefont {Lesanovsky}},\ and\ \bibinfo
  {author} {\bibfnamefont {P.}~\bibnamefont {Zoller}},\ }\bibfield  {title}
  {\bibinfo {title} {Strongly {{Correlated Gases}} of {{Rydberg}}-{{Dressed
  Atoms}}: {{Quantum}} and {{Classical Dynamics}}},\ }\href
  {https://doi.org/10.1103/PhysRevLett.104.223002} {\bibfield  {journal}
  {\bibinfo  {journal} {Phys. Rev. Lett.}\ }\textbf {\bibinfo {volume} {104}},\
  \bibinfo {pages} {223002} (\bibinfo {year} {2010})}\BibitemShut {NoStop}%
\bibitem [{\citenamefont {Balewski}\ \emph {et~al.}(2014)\citenamefont
  {Balewski}, \citenamefont {Krupp}, \citenamefont {Gaj}, \citenamefont
  {Hofferberth}, \citenamefont {Löw},\ and\ \citenamefont
  {Pfau}}]{balewski_rydberg_2014}%
  \BibitemOpen
  \bibfield  {author} {\bibinfo {author} {\bibfnamefont {J.~B.}\ \bibnamefont
  {Balewski}}, \bibinfo {author} {\bibfnamefont {A.~T.}\ \bibnamefont {Krupp}},
  \bibinfo {author} {\bibfnamefont {A.}~\bibnamefont {Gaj}}, \bibinfo {author}
  {\bibfnamefont {S.}~\bibnamefont {Hofferberth}}, \bibinfo {author}
  {\bibfnamefont {R.}~\bibnamefont {Löw}},\ and\ \bibinfo {author}
  {\bibfnamefont {T.}~\bibnamefont {Pfau}},\ }\bibfield  {title} {\bibinfo
  {title} {Rydberg dressing: Understanding of collective many-body effects and
  implications for experiments},\ }\href
  {https://doi.org/10.1088/1367-2630/16/6/063012} {\bibfield  {journal}
  {\bibinfo  {journal} {New J. Phys.}\ }\textbf {\bibinfo {volume} {16}},\
  \bibinfo {pages} {063012} (\bibinfo {year} {2014})}\BibitemShut {NoStop}%
\bibitem [{\citenamefont {Löw}\ \emph {et~al.}(2012)\citenamefont {Löw},
  \citenamefont {Weimer}, \citenamefont {Nipper}, \citenamefont {Balewski},
  \citenamefont {Butscher}, \citenamefont {Büchler},\ and\ \citenamefont
  {Pfau}}]{low_experimental_2012}%
  \BibitemOpen
  \bibfield  {author} {\bibinfo {author} {\bibfnamefont {R.}~\bibnamefont
  {Löw}}, \bibinfo {author} {\bibfnamefont {H.}~\bibnamefont {Weimer}},
  \bibinfo {author} {\bibfnamefont {J.}~\bibnamefont {Nipper}}, \bibinfo
  {author} {\bibfnamefont {J.~B.}\ \bibnamefont {Balewski}}, \bibinfo {author}
  {\bibfnamefont {B.}~\bibnamefont {Butscher}}, \bibinfo {author}
  {\bibfnamefont {H.~P.}\ \bibnamefont {Büchler}},\ and\ \bibinfo {author}
  {\bibfnamefont {T.}~\bibnamefont {Pfau}},\ }\bibfield  {title} {\bibinfo
  {title} {An experimental and theoretical guide to strongly interacting
  {{Rydberg}} gases},\ }\href {https://doi.org/10.1088/0953-4075/45/11/113001}
  {\bibfield  {journal} {\bibinfo  {journal} {J. Phys. B: At. Mol. Opt. Phys.}\
  }\textbf {\bibinfo {volume} {45}},\ \bibinfo {pages} {113001} (\bibinfo
  {year} {2012})}\BibitemShut {NoStop}%
\bibitem [{\citenamefont {Jau}\ \emph {et~al.}(2016)\citenamefont {Jau},
  \citenamefont {Hankin}, \citenamefont {Keating}, \citenamefont {Deutsch},\
  and\ \citenamefont {Biedermann}}]{Jau_Entanglingatomicspins_2016}%
  \BibitemOpen
  \bibfield  {author} {\bibinfo {author} {\bibfnamefont {Y.-Y.}\ \bibnamefont
  {Jau}}, \bibinfo {author} {\bibfnamefont {A.~M.}\ \bibnamefont {Hankin}},
  \bibinfo {author} {\bibfnamefont {T.}~\bibnamefont {Keating}}, \bibinfo
  {author} {\bibfnamefont {I.~H.}\ \bibnamefont {Deutsch}},\ and\ \bibinfo
  {author} {\bibfnamefont {G.~W.}\ \bibnamefont {Biedermann}},\ }\bibfield
  {title} {\bibinfo {title} {Entangling atomic spins with a {{Rydberg}}-dressed
  spin-flip blockade},\ }\href {https://doi.org/10.1038/nphys3487} {\bibfield
  {journal} {\bibinfo  {journal} {Nat. Phys.}\ }\textbf {\bibinfo {volume}
  {12}},\ \bibinfo {pages} {71} (\bibinfo {year} {2016})}\BibitemShut {NoStop}%
\bibitem [{\citenamefont {Zeiher}\ \emph {et~al.}(2016)\citenamefont {Zeiher},
  \citenamefont {{van Bijnen}}, \citenamefont {Schauß}, \citenamefont {Hild},
  \citenamefont {Choi}, \citenamefont {Pohl}, \citenamefont {Bloch},\ and\
  \citenamefont {Gross}}]{zeiher_manybody_2016}%
  \BibitemOpen
  \bibfield  {author} {\bibinfo {author} {\bibfnamefont {J.}~\bibnamefont
  {Zeiher}}, \bibinfo {author} {\bibfnamefont {R.}~\bibnamefont {{van
  Bijnen}}}, \bibinfo {author} {\bibfnamefont {P.}~\bibnamefont {Schauß}},
  \bibinfo {author} {\bibfnamefont {S.}~\bibnamefont {Hild}}, \bibinfo {author}
  {\bibfnamefont {J.-Y.}\ \bibnamefont {Choi}}, \bibinfo {author}
  {\bibfnamefont {T.}~\bibnamefont {Pohl}}, \bibinfo {author} {\bibfnamefont
  {I.}~\bibnamefont {Bloch}},\ and\ \bibinfo {author} {\bibfnamefont
  {C.}~\bibnamefont {Gross}},\ }\bibfield  {title} {\bibinfo {title} {Many-body
  interferometry of a {{Rydberg}}-dressed spin lattice},\ }\href
  {https://doi.org/10.1038/nphys3835} {\bibfield  {journal} {\bibinfo
  {journal} {Nat. Phys.}\ }\textbf {\bibinfo {volume} {12}},\ \bibinfo {pages}
  {1095} (\bibinfo {year} {2016})}\BibitemShut {NoStop}%
\bibitem [{\citenamefont {Zeiher}\ \emph {et~al.}(2017)\citenamefont {Zeiher},
  \citenamefont {Choi}, \citenamefont {{Rubio-Abadal}}, \citenamefont {Pohl},
  \citenamefont {{van Bijnen}}, \citenamefont {Bloch},\ and\ \citenamefont
  {Gross}}]{Zeiher_CoherentManyBodySpin_2017}%
  \BibitemOpen
  \bibfield  {author} {\bibinfo {author} {\bibfnamefont {J.}~\bibnamefont
  {Zeiher}}, \bibinfo {author} {\bibfnamefont {J.-y.}\ \bibnamefont {Choi}},
  \bibinfo {author} {\bibfnamefont {A.}~\bibnamefont {{Rubio-Abadal}}},
  \bibinfo {author} {\bibfnamefont {T.}~\bibnamefont {Pohl}}, \bibinfo {author}
  {\bibfnamefont {R.}~\bibnamefont {{van Bijnen}}}, \bibinfo {author}
  {\bibfnamefont {I.}~\bibnamefont {Bloch}},\ and\ \bibinfo {author}
  {\bibfnamefont {C.}~\bibnamefont {Gross}},\ }\bibfield  {title} {\bibinfo
  {title} {Coherent {{Many}}-{{Body Spin Dynamics}} in a {{Long}}-{{Range
  Interacting Ising Chain}}},\ }\href
  {https://doi.org/10.1103/PhysRevX.7.041063} {\bibfield  {journal} {\bibinfo
  {journal} {Phys. Rev. X}\ }\textbf {\bibinfo {volume} {7}},\ \bibinfo {pages}
  {041063} (\bibinfo {year} {2017})}\BibitemShut {NoStop}%
\bibitem [{\citenamefont {Borish}\ \emph {et~al.}(2020)\citenamefont {Borish},
  \citenamefont {Markovi{\'c}}, \citenamefont {Hines}, \citenamefont
  {Rajagopal},\ and\ \citenamefont
  {{Schleier-Smith}}}]{Borish_TransverseFieldIsingDynamics_2020}%
  \BibitemOpen
  \bibfield  {author} {\bibinfo {author} {\bibfnamefont {V.}~\bibnamefont
  {Borish}}, \bibinfo {author} {\bibfnamefont {O.}~\bibnamefont
  {Markovi{\'c}}}, \bibinfo {author} {\bibfnamefont {J.~A.}\ \bibnamefont
  {Hines}}, \bibinfo {author} {\bibfnamefont {S.~V.}\ \bibnamefont
  {Rajagopal}},\ and\ \bibinfo {author} {\bibfnamefont {M.}~\bibnamefont
  {{Schleier-Smith}}},\ }\bibfield  {title} {\bibinfo {title}
  {Transverse-{{Field Ising Dynamics}} in a {{Rydberg}}-{{Dressed Atomic
  Gas}}},\ }\href {https://doi.org/10.1103/PhysRevLett.124.063601} {\bibfield
  {journal} {\bibinfo  {journal} {Phys. Rev. Lett.}\ }\textbf {\bibinfo
  {volume} {124}},\ \bibinfo {pages} {063601} (\bibinfo {year}
  {2020})}\BibitemShut {NoStop}%
\bibitem [{\citenamefont {{Guardado-Sanchez}}\ \emph
  {et~al.}(2021)\citenamefont {{Guardado-Sanchez}}, \citenamefont {Spar},
  \citenamefont {Schauss}, \citenamefont {Belyansky}, \citenamefont {Young},
  \citenamefont {Bienias}, \citenamefont {Gorshkov}, \citenamefont {Iadecola},\
  and\ \citenamefont {Bakr}}]{Guardado-Sanchez_QuenchDynamicsFermi_2021}%
  \BibitemOpen
  \bibfield  {author} {\bibinfo {author} {\bibfnamefont {E.}~\bibnamefont
  {{Guardado-Sanchez}}}, \bibinfo {author} {\bibfnamefont {B.~M.}\ \bibnamefont
  {Spar}}, \bibinfo {author} {\bibfnamefont {P.}~\bibnamefont {Schauss}},
  \bibinfo {author} {\bibfnamefont {R.}~\bibnamefont {Belyansky}}, \bibinfo
  {author} {\bibfnamefont {J.~T.}\ \bibnamefont {Young}}, \bibinfo {author}
  {\bibfnamefont {P.}~\bibnamefont {Bienias}}, \bibinfo {author} {\bibfnamefont
  {A.~V.}\ \bibnamefont {Gorshkov}}, \bibinfo {author} {\bibfnamefont
  {T.}~\bibnamefont {Iadecola}},\ and\ \bibinfo {author} {\bibfnamefont
  {W.~S.}\ \bibnamefont {Bakr}},\ }\bibfield  {title} {\bibinfo {title} {Quench
  {{Dynamics}} of a {{Fermi Gas}} with {{Strong Nonlocal Interactions}}},\
  }\href {https://doi.org/10.1103/PhysRevX.11.021036} {\bibfield  {journal}
  {\bibinfo  {journal} {Phys. Rev. X}\ }\textbf {\bibinfo {volume} {11}},\
  \bibinfo {pages} {021036} (\bibinfo {year} {2021})}\BibitemShut {NoStop}%
\bibitem [{\citenamefont {Macr{\`i}}\ \emph {et~al.}(2014)\citenamefont
  {Macr{\`i}}, \citenamefont {Saccani},\ and\ \citenamefont
  {Cinti}}]{Macri_GroundStateExcitation_2014}%
  \BibitemOpen
  \bibfield  {author} {\bibinfo {author} {\bibfnamefont {T.}~\bibnamefont
  {Macr{\`i}}}, \bibinfo {author} {\bibfnamefont {S.}~\bibnamefont {Saccani}},\
  and\ \bibinfo {author} {\bibfnamefont {F.}~\bibnamefont {Cinti}},\ }\bibfield
   {title} {\bibinfo {title} {Ground {{State}} and {{Excitation Properties}} of
  {{Soft}}-{{Core Bosons}}},\ }\href
  {https://doi.org/10.1007/s10909-014-1192-7} {\bibfield  {journal} {\bibinfo
  {journal} {J. Low Temp. Phys.}\ }\textbf {\bibinfo {volume} {177}},\ \bibinfo
  {pages} {59} (\bibinfo {year} {2014})}\BibitemShut {NoStop}%
\bibitem [{\citenamefont {Prestipino}\ \emph {et~al.}(2019)\citenamefont
  {Prestipino}, \citenamefont {Sergi},\ and\ \citenamefont
  {Bruno}}]{Prestipino_Clusterizationweaklyinteractingbosons_2019}%
  \BibitemOpen
  \bibfield  {author} {\bibinfo {author} {\bibfnamefont {S.}~\bibnamefont
  {Prestipino}}, \bibinfo {author} {\bibfnamefont {A.}~\bibnamefont {Sergi}},\
  and\ \bibinfo {author} {\bibfnamefont {E.}~\bibnamefont {Bruno}},\ }\bibfield
   {title} {\bibinfo {title} {Clusterization of weakly-interacting bosons in
  one dimension: An analytic study at zero temperature},\ }\href
  {https://doi.org/10.1088/1751-8121/aaee94} {\bibfield  {journal} {\bibinfo
  {journal} {J. Phys. A: Math. Theor.}\ }\textbf {\bibinfo {volume} {52}},\
  \bibinfo {pages} {015002} (\bibinfo {year} {2019})}\BibitemShut {NoStop}%
\bibitem [{\citenamefont {Lieb}\ and\ \citenamefont
  {Liniger}(1963)}]{lieb_exact_1963a}%
  \BibitemOpen
  \bibfield  {author} {\bibinfo {author} {\bibfnamefont {E.~H.}\ \bibnamefont
  {Lieb}}\ and\ \bibinfo {author} {\bibfnamefont {W.}~\bibnamefont {Liniger}},\
  }\bibfield  {title} {\bibinfo {title} {Exact {{Analysis}} of an {{Interacting
  Bose Gas}}. {{I}}. {{The General Solution}} and the {{Ground State}}},\
  }\href {https://doi.org/10.1103/PhysRev.130.1605} {\bibfield  {journal}
  {\bibinfo  {journal} {Phys. Rev.}\ }\textbf {\bibinfo {volume} {130}},\
  \bibinfo {pages} {1605} (\bibinfo {year} {1963})}\BibitemShut {NoStop}%
\bibitem [{\citenamefont {Teruzzi}\ \emph {et~al.}(2017)\citenamefont
  {Teruzzi}, \citenamefont {Galli},\ and\ \citenamefont {Bertaina}}]{QFS}%
  \BibitemOpen
  \bibfield  {author} {\bibinfo {author} {\bibfnamefont {M.}~\bibnamefont
  {Teruzzi}}, \bibinfo {author} {\bibfnamefont {D.~E.}\ \bibnamefont {Galli}},\
  and\ \bibinfo {author} {\bibfnamefont {G.}~\bibnamefont {Bertaina}},\
  }\bibfield  {title} {\bibinfo {title} {Microscopic study of static and
  dynamical properties of dilute one--dimensional soft bosons},\ }\href
  {https://doi.org/10.1007/s10909-016-1736-0} {\bibfield  {journal} {\bibinfo
  {journal} {{J.} {Low} {Temp.} {Phys.}}\ }\textbf {\bibinfo {volume} {187}},\
  \bibinfo {pages} {719} (\bibinfo {year} {2017})}\BibitemShut {NoStop}%
\bibitem [{\citenamefont {Tonks}(1936)}]{Tonks}%
  \BibitemOpen
  \bibfield  {author} {\bibinfo {author} {\bibfnamefont {L.}~\bibnamefont
  {Tonks}},\ }\bibfield  {title} {\bibinfo {title} {The complete equation of
  state of one, two and three--dimensional gases of hard elastic spheres},\
  }\href {https://doi.org/10.1103/PhysRev.50.955} {\bibfield  {journal}
  {\bibinfo  {journal} {{Phys}. {Rev}.}\ }\textbf {\bibinfo {volume} {50}},\
  \bibinfo {pages} {955} (\bibinfo {year} {1936})}\BibitemShut {NoStop}%
\bibitem [{\citenamefont {Girardeau}(1960)}]{Girardeau}%
  \BibitemOpen
  \bibfield  {author} {\bibinfo {author} {\bibfnamefont {M.}~\bibnamefont
  {Girardeau}},\ }\bibfield  {title} {\bibinfo {title} {Relationship between
  systems of impenetrable bosons and fermions in one dimension},\ }\href@noop
  {} {\bibfield  {journal} {\bibinfo  {journal} {{J.} {Math.} {Phys.}}\
  }\textbf {\bibinfo {volume} {1}},\ \bibinfo {pages} {516} (\bibinfo {year}
  {1960})}\BibitemShut {NoStop}%
\bibitem [{\citenamefont {Olshanii}(1998)}]{olshanii1998atomic}%
  \BibitemOpen
  \bibfield  {author} {\bibinfo {author} {\bibfnamefont {M.}~\bibnamefont
  {Olshanii}},\ }\bibfield  {title} {\bibinfo {title} {Atomic scattering in the
  presence of an external confinement and a gas of impenetrable bosons},\
  }\href@noop {} {\bibfield  {journal} {\bibinfo  {journal} {{Phys.} {Rev.}
  {Lett.}}\ }\textbf {\bibinfo {volume} {81}},\ \bibinfo {pages} {938}
  (\bibinfo {year} {1998})}\BibitemShut {NoStop}%
\bibitem [{\citenamefont {P{\l}odzie{\'n}}\ \emph {et~al.}(2017)\citenamefont
  {P{\l}odzie{\'n}}, \citenamefont {Lochead}, \citenamefont {{de Hond}},
  \citenamefont {{van Druten}},\ and\ \citenamefont
  {Kokkelmans}}]{plodzien_rydberg_2017}%
  \BibitemOpen
  \bibfield  {author} {\bibinfo {author} {\bibfnamefont {M.}~\bibnamefont
  {P{\l}odzie{\'n}}}, \bibinfo {author} {\bibfnamefont {G.}~\bibnamefont
  {Lochead}}, \bibinfo {author} {\bibfnamefont {J.}~\bibnamefont {{de Hond}}},
  \bibinfo {author} {\bibfnamefont {N.~J.}\ \bibnamefont {{van Druten}}},\ and\
  \bibinfo {author} {\bibfnamefont {S.}~\bibnamefont {Kokkelmans}},\ }\bibfield
   {title} {\bibinfo {title} {Rydberg dressing of a one-dimensional
  {{Bose}}-{{Einstein}} condensate},\ }\href
  {https://doi.org/10.1103/PhysRevA.95.043606} {\bibfield  {journal} {\bibinfo
  {journal} {Phys. Rev. A}\ }\textbf {\bibinfo {volume} {95}},\ \bibinfo
  {pages} {043606} (\bibinfo {year} {2017})}\BibitemShut {NoStop}%
\bibitem [{\citenamefont {Teruzzi}\ \emph {et~al.}(2018)\citenamefont
  {Teruzzi}, \citenamefont {Pini}, \citenamefont {Rossotti}, \citenamefont
  {Galli},\ and\ \citenamefont {Bertaina}}]{StaticKorea}%
  \BibitemOpen
  \bibfield  {author} {\bibinfo {author} {\bibfnamefont {M.}~\bibnamefont
  {Teruzzi}}, \bibinfo {author} {\bibfnamefont {D.}~\bibnamefont {Pini}},
  \bibinfo {author} {\bibfnamefont {S.}~\bibnamefont {Rossotti}}, \bibinfo
  {author} {\bibfnamefont {D.~E.}\ \bibnamefont {Galli}},\ and\ \bibinfo
  {author} {\bibfnamefont {G.}~\bibnamefont {Bertaina}},\ }\bibfield  {title}
  {\bibinfo {title} {Static density response of one-dimensional soft bosons
  across the clustering transition},\ }\href
  {http://stacks.iop.org/1742-6596/1041/i=1/a=012009} {\bibfield  {journal}
  {\bibinfo  {journal} {J. Phys.: Conf. Series}\ }\textbf {\bibinfo {volume}
  {1041}},\ \bibinfo {pages} {012009} (\bibinfo {year} {2018})}\BibitemShut
  {NoStop}%
\bibitem [{\citenamefont {Prestipino}(2014)}]{prestipino_cluster_2014}%
  \BibitemOpen
  \bibfield  {author} {\bibinfo {author} {\bibfnamefont {S.}~\bibnamefont
  {Prestipino}},\ }\bibfield  {title} {\bibinfo {title} {Cluster phases of
  penetrable rods on a line},\ }\href
  {https://doi.org/10.1103/PhysRevE.90.042306} {\bibfield  {journal} {\bibinfo
  {journal} {Phys. Rev. E}\ }\textbf {\bibinfo {volume} {90}},\ \bibinfo
  {pages} {042306} (\bibinfo {year} {2014})}\BibitemShut {NoStop}%
\bibitem [{\citenamefont {Prestipino}\ \emph {et~al.}(2015)\citenamefont
  {Prestipino}, \citenamefont {Gazzillo},\ and\ \citenamefont
  {Tasinato}}]{prestipino_probing_2015}%
  \BibitemOpen
  \bibfield  {author} {\bibinfo {author} {\bibfnamefont {S.}~\bibnamefont
  {Prestipino}}, \bibinfo {author} {\bibfnamefont {D.}~\bibnamefont
  {Gazzillo}},\ and\ \bibinfo {author} {\bibfnamefont {N.}~\bibnamefont
  {Tasinato}},\ }\bibfield  {title} {\bibinfo {title} {Probing the existence of
  phase transitions in one-dimensional fluids of penetrable particles},\ }\href
  {https://doi.org/10.1103/PhysRevE.92.022138} {\bibfield  {journal} {\bibinfo
  {journal} {Physical Review E}\ }\textbf {\bibinfo {volume} {92}},\ \bibinfo
  {pages} {022138} (\bibinfo {year} {2015})}\BibitemShut {NoStop}%
\bibitem [{\citenamefont {Mambretti}\ \emph {et~al.}(2020)\citenamefont
  {Mambretti}, \citenamefont {Molinelli}, \citenamefont {Pini}, \citenamefont
  {Bertaina},\ and\ \citenamefont
  {Galli}}]{Mambretti_EmergenceIsingcritical_2020}%
  \BibitemOpen
  \bibfield  {author} {\bibinfo {author} {\bibfnamefont {F.}~\bibnamefont
  {Mambretti}}, \bibinfo {author} {\bibfnamefont {S.}~\bibnamefont
  {Molinelli}}, \bibinfo {author} {\bibfnamefont {D.}~\bibnamefont {Pini}},
  \bibinfo {author} {\bibfnamefont {G.}~\bibnamefont {Bertaina}},\ and\
  \bibinfo {author} {\bibfnamefont {D.~E.}\ \bibnamefont {Galli}},\ }\bibfield
  {title} {\bibinfo {title} {Emergence of an {{Ising}} critical regime in the
  clustering of one-dimensional soft matter revealed through string
  variables},\ }\href {https://doi.org/10.1103/PhysRevE.102.042134} {\bibfield
  {journal} {\bibinfo  {journal} {Phys. Rev. E}\ }\textbf {\bibinfo {volume}
  {102}},\ \bibinfo {pages} {042134} (\bibinfo {year} {2020})}\BibitemShut
  {NoStop}%
\bibitem [{\citenamefont {Caux}\ and\ \citenamefont
  {Calabrese}(2006)}]{caux_dynamical_2006}%
  \BibitemOpen
  \bibfield  {author} {\bibinfo {author} {\bibfnamefont {J.-S.}\ \bibnamefont
  {Caux}}\ and\ \bibinfo {author} {\bibfnamefont {P.}~\bibnamefont
  {Calabrese}},\ }\bibfield  {title} {\bibinfo {title} {Dynamical
  density-density correlations in the one-dimensional {{Bose}} gas},\ }\href
  {https://doi.org/10.1103/PhysRevA.74.031605} {\bibfield  {journal} {\bibinfo
  {journal} {Phys. Rev. A}\ }\textbf {\bibinfo {volume} {74}},\ \bibinfo
  {pages} {031605} (\bibinfo {year} {2006})}\BibitemShut {NoStop}%
\bibitem [{\citenamefont {Motta}\ \emph {et~al.}(2016)\citenamefont {Motta},
  \citenamefont {Vitali}, \citenamefont {Rossi}, \citenamefont {Galli},\ and\
  \citenamefont {Bertaina}}]{motta_dynamical_2016}%
  \BibitemOpen
  \bibfield  {author} {\bibinfo {author} {\bibfnamefont {M.}~\bibnamefont
  {Motta}}, \bibinfo {author} {\bibfnamefont {E.}~\bibnamefont {Vitali}},
  \bibinfo {author} {\bibfnamefont {M.}~\bibnamefont {Rossi}}, \bibinfo
  {author} {\bibfnamefont {D.~E.}\ \bibnamefont {Galli}},\ and\ \bibinfo
  {author} {\bibfnamefont {G.}~\bibnamefont {Bertaina}},\ }\bibfield  {title}
  {\bibinfo {title} {Dynamical structure factor of one-dimensional hard rods},\
  }\href {https://doi.org/10.1103/PhysRevA.94.043627} {\bibfield  {journal}
  {\bibinfo  {journal} {Phys. Rev. A}\ }\textbf {\bibinfo {volume} {94}},\
  \bibinfo {pages} {043627} (\bibinfo {year} {2016})}\BibitemShut {NoStop}%
\bibitem [{\citenamefont {McMillan}(1965)}]{McMillan_GroundStateLiquid_1965}%
  \BibitemOpen
  \bibfield  {author} {\bibinfo {author} {\bibfnamefont {W.~L.}\ \bibnamefont
  {McMillan}},\ }\bibfield  {title} {\bibinfo {title} {Ground {{State}} of
  {{Liquid}} {{He}}$^4$},\ }\href {https://doi.org/10.1103/PhysRev.138.A442}
  {\bibfield  {journal} {\bibinfo  {journal} {Phys. Rev.}\ }\textbf {\bibinfo
  {volume} {138}},\ \bibinfo {pages} {A442} (\bibinfo {year}
  {1965})}\BibitemShut {NoStop}%
\bibitem [{\citenamefont {Sarsa}\ \emph {et~al.}(2000)\citenamefont {Sarsa},
  \citenamefont {Schmidt},\ and\ \citenamefont {Magro}}]{PIGS}%
  \BibitemOpen
  \bibfield  {author} {\bibinfo {author} {\bibfnamefont {A.}~\bibnamefont
  {Sarsa}}, \bibinfo {author} {\bibfnamefont {K.~E.}\ \bibnamefont {Schmidt}},\
  and\ \bibinfo {author} {\bibfnamefont {W.~R.}\ \bibnamefont {Magro}},\
  }\bibfield  {title} {\bibinfo {title} {A path integral ground state method},\
  }\href {https://doi.org/http://dx.doi.org/10.1063/1.481926} {\bibfield
  {journal} {\bibinfo  {journal} {{J.} {Chem.} {Phys.}}\ }\textbf {\bibinfo
  {volume} {113}},\ \bibinfo {pages} {1366} (\bibinfo {year}
  {2000})}\BibitemShut {NoStop}%
\bibitem [{\citenamefont {Galli}\ and\ \citenamefont {Reatto}(2003)}]{SPIGS}%
  \BibitemOpen
  \bibfield  {author} {\bibinfo {author} {\bibfnamefont {D.~E.}\ \bibnamefont
  {Galli}}\ and\ \bibinfo {author} {\bibfnamefont {L.}~\bibnamefont {Reatto}},\
  }\bibfield  {title} {\bibinfo {title} {Recent progress in simulation of the
  ground state of many boson systems},\ }\href
  {https://doi.org/10.1080/0026897031000074562} {\bibfield  {journal} {\bibinfo
   {journal} {{Mol.} {Phys.}}\ }\textbf {\bibinfo {volume} {101}},\ \bibinfo
  {pages} {1697} (\bibinfo {year} {2003})}\BibitemShut {NoStop}%
\bibitem [{\citenamefont {Rossi}\ \emph {et~al.}(2009)\citenamefont {Rossi},
  \citenamefont {Nava}, \citenamefont {Reatto},\ and\ \citenamefont
  {Galli}}]{Rossi_Exactgroundstate_2009}%
  \BibitemOpen
  \bibfield  {author} {\bibinfo {author} {\bibfnamefont {M.}~\bibnamefont
  {Rossi}}, \bibinfo {author} {\bibfnamefont {M.}~\bibnamefont {Nava}},
  \bibinfo {author} {\bibfnamefont {L.}~\bibnamefont {Reatto}},\ and\ \bibinfo
  {author} {\bibfnamefont {D.~E.}\ \bibnamefont {Galli}},\ }\bibfield  {title}
  {\bibinfo {title} {Exact ground state {{Monte Carlo}} method for {{Bosons}}
  without importance sampling},\ }\href {https://doi.org/10.1063/1.3247833}
  {\bibfield  {journal} {\bibinfo  {journal} {J. Chem. Phys.}\ }\textbf
  {\bibinfo {volume} {131}},\ \bibinfo {pages} {154108} (\bibinfo {year}
  {2009})}\BibitemShut {NoStop}%
\bibitem [{\citenamefont {Reatto}\ and\ \citenamefont
  {Chester}(1967)}]{ReattoChester}%
  \BibitemOpen
  \bibfield  {author} {\bibinfo {author} {\bibfnamefont {L.}~\bibnamefont
  {Reatto}}\ and\ \bibinfo {author} {\bibfnamefont {G.~V.}\ \bibnamefont
  {Chester}},\ }\bibfield  {title} {\bibinfo {title} {Phonons and the
  properties of a {Bose} system},\ }\href
  {https://doi.org/10.1103/PhysRev.155.88} {\bibfield  {journal} {\bibinfo
  {journal} {{Phys.} {Rev.}}\ }\textbf {\bibinfo {volume} {155}},\ \bibinfo
  {pages} {88} (\bibinfo {year} {1967})}\BibitemShut {NoStop}%
\bibitem [{\citenamefont {Vitali}\ \emph {et~al.}(2010)\citenamefont {Vitali},
  \citenamefont {Rossi}, \citenamefont {Reatto},\ and\ \citenamefont
  {Galli}}]{GIFT}%
  \BibitemOpen
  \bibfield  {author} {\bibinfo {author} {\bibfnamefont {E.}~\bibnamefont
  {Vitali}}, \bibinfo {author} {\bibfnamefont {M.}~\bibnamefont {Rossi}},
  \bibinfo {author} {\bibfnamefont {L.}~\bibnamefont {Reatto}},\ and\ \bibinfo
  {author} {\bibfnamefont {D.~E.}\ \bibnamefont {Galli}},\ }\bibfield  {title}
  {\bibinfo {title} {Ab initio low--energy dynamics of superfluid and solid
  $^{4}\mathrm{{H}e}$},\ }\href@noop {} {\bibfield  {journal} {\bibinfo
  {journal} {{Phys.} {Rev.} B}\ }\textbf {\bibinfo {volume} {82}},\ \bibinfo
  {pages} {174510} (\bibinfo {year} {2010})}\BibitemShut {NoStop}%
\bibitem [{\citenamefont {Bertaina}\ \emph {et~al.}(2016)\citenamefont
  {Bertaina}, \citenamefont {Motta}, \citenamefont {Rossi}, \citenamefont
  {Vitali},\ and\ \citenamefont {Galli}}]{bertaina_onedimensional_2016}%
  \BibitemOpen
  \bibfield  {author} {\bibinfo {author} {\bibfnamefont {G.}~\bibnamefont
  {Bertaina}}, \bibinfo {author} {\bibfnamefont {M.}~\bibnamefont {Motta}},
  \bibinfo {author} {\bibfnamefont {M.}~\bibnamefont {Rossi}}, \bibinfo
  {author} {\bibfnamefont {E.}~\bibnamefont {Vitali}},\ and\ \bibinfo {author}
  {\bibfnamefont {D.~E.}\ \bibnamefont {Galli}},\ }\bibfield  {title} {\bibinfo
  {title} {One-dimensional liquid $^{4}\mathrm{He}$: Dynamical properties
  beyond luttinger-liquid theory},\ }\href
  {https://doi.org/10.1103/PhysRevLett.116.135302} {\bibfield  {journal}
  {\bibinfo  {journal} {Phys. Rev. Lett.}\ }\textbf {\bibinfo {volume} {116}},\
  \bibinfo {pages} {135302} (\bibinfo {year} {2016})}\BibitemShut {NoStop}%
\bibitem [{\citenamefont {Bertaina}\ \emph {et~al.}(2017)\citenamefont
  {Bertaina}, \citenamefont {Galli},\ and\ \citenamefont
  {Vitali}}]{Bertaina_Statisticalcomputationalintelligence_2017}%
  \BibitemOpen
  \bibfield  {author} {\bibinfo {author} {\bibfnamefont {G.}~\bibnamefont
  {Bertaina}}, \bibinfo {author} {\bibfnamefont {D.~E.}\ \bibnamefont
  {Galli}},\ and\ \bibinfo {author} {\bibfnamefont {E.}~\bibnamefont
  {Vitali}},\ }\bibfield  {title} {\bibinfo {title} {Statistical and
  computational intelligence approach to analytic continuation in {{Quantum
  Monte Carlo}}},\ }\href {https://doi.org/10.1080/23746149.2017.1288585}
  {\bibfield  {journal} {\bibinfo  {journal} {Adv. Phys.: X}\ }\textbf
  {\bibinfo {volume} {2}},\ \bibinfo {pages} {302} (\bibinfo {year}
  {2017})}\BibitemShut {NoStop}%
\bibitem [{\citenamefont {Meyer}\ \emph {et~al.}(1990)\citenamefont {Meyer},
  \citenamefont {Manthe},\ and\ \citenamefont {Cederbaum}}]{Meyer_MCTDH}%
  \BibitemOpen
  \bibfield  {author} {\bibinfo {author} {\bibfnamefont {H.-D.}\ \bibnamefont
  {Meyer}}, \bibinfo {author} {\bibfnamefont {U.}~\bibnamefont {Manthe}},\ and\
  \bibinfo {author} {\bibfnamefont {L.}~\bibnamefont {Cederbaum}},\ }\bibfield
  {title} {\bibinfo {title} {The multi-configurational time-dependent {H}artree
  approach},\ }\href
  {https://doi.org/https://doi.org/10.1016/0009-2614(90)87014-I} {\bibfield
  {journal} {\bibinfo  {journal} {Chem. Phys. Lett.}\ }\textbf {\bibinfo
  {volume} {165}},\ \bibinfo {pages} {73} (\bibinfo {year} {1990})}\BibitemShut
  {NoStop}%
\bibitem [{\citenamefont {Alon}\ \emph {et~al.}(2007)\citenamefont {Alon},
  \citenamefont {Streltsov},\ and\ \citenamefont {Cederbaum}}]{Alon_MCTDH}%
  \BibitemOpen
  \bibfield  {author} {\bibinfo {author} {\bibfnamefont {O.~E.}\ \bibnamefont
  {Alon}}, \bibinfo {author} {\bibfnamefont {A.~I.}\ \bibnamefont
  {Streltsov}},\ and\ \bibinfo {author} {\bibfnamefont {L.~S.}\ \bibnamefont
  {Cederbaum}},\ }\bibfield  {title} {\bibinfo {title} {Unified view on
  multiconfigurational time propagation for systems consisting of identical
  particles},\ }\href {https://doi.org/10.1063/1.2771159} {\bibfield  {journal}
  {\bibinfo  {journal} {J. Chem. Phys.}\ }\textbf {\bibinfo {volume} {127}},\
  \bibinfo {pages} {154103} (\bibinfo {year} {2007})}\BibitemShut {NoStop}%
\bibitem [{\citenamefont {Lode}\ \emph {et~al.}(2020)\citenamefont {Lode},
  \citenamefont {Tsatsos}, \citenamefont {Fasshauer}, \citenamefont {Lin},
  \citenamefont {Papariello}, \citenamefont {Molignini}, \citenamefont
  {L{\'{e}}v{\^{e}}que},\ and\ \citenamefont {Weiner}}]{MCTDH-X}%
  \BibitemOpen
  \bibfield  {author} {\bibinfo {author} {\bibfnamefont {A.~U.~J.}\
  \bibnamefont {Lode}}, \bibinfo {author} {\bibfnamefont {M.~C.}\ \bibnamefont
  {Tsatsos}}, \bibinfo {author} {\bibfnamefont {E.}~\bibnamefont {Fasshauer}},
  \bibinfo {author} {\bibfnamefont {R.}~\bibnamefont {Lin}}, \bibinfo {author}
  {\bibfnamefont {L.}~\bibnamefont {Papariello}}, \bibinfo {author}
  {\bibfnamefont {P.}~\bibnamefont {Molignini}}, \bibinfo {author}
  {\bibfnamefont {C.}~\bibnamefont {L{\'{e}}v{\^{e}}que}},\ and\ \bibinfo
  {author} {\bibfnamefont {S.~E.}\ \bibnamefont {Weiner}},\ }\href@noop {}
  {\bibinfo {title} {{MCTDH-X}: The time-dependent multiconfigurational
  {H}artree for indistinguishable particles software,
  \url{http://ultracold.org}}} (\bibinfo {year} {2020})\BibitemShut {NoStop}%
\bibitem [{\citenamefont {Lin}\ \emph {et~al.}(2020)\citenamefont {Lin},
  \citenamefont {Molignini}, \citenamefont {Papariello}, \citenamefont
  {Tsatsos}, \citenamefont {L{\'{e}}v{\^{e}}que}, \citenamefont {Weiner},
  \citenamefont {Fasshauer}, \citenamefont {Chitra},\ and\ \citenamefont
  {Lode}}]{Lin_2020}%
  \BibitemOpen
  \bibfield  {author} {\bibinfo {author} {\bibfnamefont {R.}~\bibnamefont
  {Lin}}, \bibinfo {author} {\bibfnamefont {P.}~\bibnamefont {Molignini}},
  \bibinfo {author} {\bibfnamefont {L.}~\bibnamefont {Papariello}}, \bibinfo
  {author} {\bibfnamefont {M.~C.}\ \bibnamefont {Tsatsos}}, \bibinfo {author}
  {\bibfnamefont {C.}~\bibnamefont {L{\'{e}}v{\^{e}}que}}, \bibinfo {author}
  {\bibfnamefont {S.~E.}\ \bibnamefont {Weiner}}, \bibinfo {author}
  {\bibfnamefont {E.}~\bibnamefont {Fasshauer}}, \bibinfo {author}
  {\bibfnamefont {R.}~\bibnamefont {Chitra}},\ and\ \bibinfo {author}
  {\bibfnamefont {A.~U.~J.}\ \bibnamefont {Lode}},\ }\bibfield  {title}
  {\bibinfo {title} {{MCTDH-X}: The multiconfigurational time-dependent
  {H}artree method for indistinguishable particles software},\ }\href
  {https://doi.org/10.1088/2058-9565/ab788b} {\bibfield  {journal} {\bibinfo
  {journal} {Quantum Sci. Technol.}\ }\textbf {\bibinfo {volume} {5}},\
  \bibinfo {pages} {024004} (\bibinfo {year} {2020})}\BibitemShut {NoStop}%
\bibitem [{\citenamefont {Lode}(2016)}]{Lode_2016}%
  \BibitemOpen
  \bibfield  {author} {\bibinfo {author} {\bibfnamefont {A.~U.~J.}\
  \bibnamefont {Lode}},\ }\bibfield  {title} {\bibinfo {title}
  {Multiconfigurational time-dependent {H}artree method for bosons with
  internal degrees of freedom: Theory and composite fragmentation of
  multicomponent {B}ose-{E}instein condensates},\ }\href
  {https://doi.org/10.1103/PhysRevA.93.063601} {\bibfield  {journal} {\bibinfo
  {journal} {Phys. Rev. A}\ }\textbf {\bibinfo {volume} {93}},\ \bibinfo
  {pages} {063601} (\bibinfo {year} {2016})}\BibitemShut {NoStop}%
\bibitem [{\citenamefont {Fasshauer}\ and\ \citenamefont
  {Lode}(2016)}]{Fasshauer_2016}%
  \BibitemOpen
  \bibfield  {author} {\bibinfo {author} {\bibfnamefont {E.}~\bibnamefont
  {Fasshauer}}\ and\ \bibinfo {author} {\bibfnamefont {A.~U.~J.}\ \bibnamefont
  {Lode}},\ }\bibfield  {title} {\bibinfo {title} {Multiconfigurational
  time-dependent {H}artree method for fermions: Implementation, exactness, and
  few-fermion tunneling to open space},\ }\href
  {https://doi.org/10.1103/PhysRevA.93.033635} {\bibfield  {journal} {\bibinfo
  {journal} {Phys. Rev. A}\ }\textbf {\bibinfo {volume} {93}},\ \bibinfo
  {pages} {033635} (\bibinfo {year} {2016})}\BibitemShut {NoStop}%
\bibitem [{\citenamefont {Kramer}\ and\ \citenamefont
  {Saraceno}(1981)}]{Kramer}%
  \BibitemOpen
  \bibfield  {author} {\bibinfo {author} {\bibfnamefont {P.}~\bibnamefont
  {Kramer}}\ and\ \bibinfo {author} {\bibfnamefont {M.}~\bibnamefont
  {Saraceno}},\ }\href {https://doi.org/10.1007/3-540-10579-4} {\emph {\bibinfo
  {title} {Geometry of the Time-Dependent Variational Principle in Quantum
  Mechanics}}}\ (\bibinfo  {publisher} {Springer-Verlag Berlin Heidelberg},\
  \bibinfo {year} {1981})\BibitemShut {NoStop}%
\bibitem [{\citenamefont {Kull}\ and\ \citenamefont {Pfirsch}(2000)}]{Kull}%
  \BibitemOpen
  \bibfield  {author} {\bibinfo {author} {\bibfnamefont {H.-J.}\ \bibnamefont
  {Kull}}\ and\ \bibinfo {author} {\bibfnamefont {D.}~\bibnamefont {Pfirsch}},\
  }\bibfield  {title} {\bibinfo {title} {Generalized variational principle for
  the time-dependent hartree-fock equations for a slater determinant},\ }\href
  {https://doi.org/10.1103/PhysRevE.61.5940} {\bibfield  {journal} {\bibinfo
  {journal} {Phys. Rev. E}\ }\textbf {\bibinfo {volume} {61}},\ \bibinfo
  {pages} {5940} (\bibinfo {year} {2000})}\BibitemShut {NoStop}%
\bibitem [{\citenamefont
  {Apostoli}(2020)}]{master_thesis_Christian_Apostoli_2020}%
  \BibitemOpen
  \bibfield  {author} {\bibinfo {author} {\bibfnamefont {C.}~\bibnamefont
  {Apostoli}},\ }\emph {\bibinfo {title} {Study of ultracold Rydberg gases via
  the multiconfiguration time-dependent approach}},\ \href@noop {} {Master's
  thesis},\ \bibinfo  {school} {Department of Physics, University of Milan}
  (\bibinfo {year} {2020})\BibitemShut {NoStop}%
\bibitem [{\citenamefont {Astrakharchik}\ \emph {et~al.}(2005)\citenamefont
  {Astrakharchik}, \citenamefont {Boronat}, \citenamefont {Casulleras},\ and\
  \citenamefont {Giorgini}}]{astrakharchik_tonks_2005}%
  \BibitemOpen
  \bibfield  {author} {\bibinfo {author} {\bibfnamefont {G.~E.}\ \bibnamefont
  {Astrakharchik}}, \bibinfo {author} {\bibfnamefont {J.}~\bibnamefont
  {Boronat}}, \bibinfo {author} {\bibfnamefont {J.}~\bibnamefont
  {Casulleras}},\ and\ \bibinfo {author} {\bibfnamefont {S.}~\bibnamefont
  {Giorgini}},\ }\bibfield  {title} {\bibinfo {title} {Beyond the
  {{Tonks}}-{{Girardeau Gas}}: {{Strongly Correlated Regime}} in
  {{Quasi}}-{{One}}-{{Dimensional Bose Gases}}},\ }\href
  {https://doi.org/10.1103/PhysRevLett.95.190407} {\bibfield  {journal}
  {\bibinfo  {journal} {Phys. Rev. Lett.}\ }\textbf {\bibinfo {volume} {95}},\
  \bibinfo {pages} {190407} (\bibinfo {year} {2005})}\BibitemShut {NoStop}%
\bibitem [{\citenamefont {Mazzanti}\ \emph {et~al.}(2008)\citenamefont
  {Mazzanti}, \citenamefont {Astrakharchik}, \citenamefont {Boronat},\ and\
  \citenamefont {Casulleras}}]{mazzanti_ground_2008}%
  \BibitemOpen
  \bibfield  {author} {\bibinfo {author} {\bibfnamefont {F.}~\bibnamefont
  {Mazzanti}}, \bibinfo {author} {\bibfnamefont {G.~E.}\ \bibnamefont
  {Astrakharchik}}, \bibinfo {author} {\bibfnamefont {J.}~\bibnamefont
  {Boronat}},\ and\ \bibinfo {author} {\bibfnamefont {J.}~\bibnamefont
  {Casulleras}},\ }\bibfield  {title} {\bibinfo {title} {Ground-{{State
  Properties}} of a {{One}}-{{Dimensional System}} of {{Hard Rods}}},\ }\href
  {https://doi.org/10.1103/PhysRevLett.100.020401} {\bibfield  {journal}
  {\bibinfo  {journal} {Phys. Rev. Lett.}\ }\textbf {\bibinfo {volume} {100}},\
  \bibinfo {pages} {020401} (\bibinfo {year} {2008})}\BibitemShut {NoStop}%
\bibitem [{\citenamefont {Lang}\ \emph {et~al.}(2000)\citenamefont {Lang},
  \citenamefont {Likos}, \citenamefont {Watzlawek},\ and\ \citenamefont
  {L{\"o}wen}}]{lang_fluid_2000}%
  \BibitemOpen
  \bibfield  {author} {\bibinfo {author} {\bibfnamefont {A.}~\bibnamefont
  {Lang}}, \bibinfo {author} {\bibfnamefont {C.~N.}\ \bibnamefont {Likos}},
  \bibinfo {author} {\bibfnamefont {M.}~\bibnamefont {Watzlawek}},\ and\
  \bibinfo {author} {\bibfnamefont {H.}~\bibnamefont {L{\"o}wen}},\ }\bibfield
  {title} {\bibinfo {title} {Fluid and solid phases of the {{Gaussian}} core
  model},\ }\href {https://doi.org/10.1088/0953-8984/12/24/302} {\bibfield
  {journal} {\bibinfo  {journal} {J. Phys.: Condens. Matter}\ }\textbf
  {\bibinfo {volume} {12}},\ \bibinfo {pages} {5087} (\bibinfo {year}
  {2000})}\BibitemShut {NoStop}%
\bibitem [{\citenamefont {Teruzzi}\ \emph {et~al.}(2021)\citenamefont
  {Teruzzi}, \citenamefont {Apostoli}, \citenamefont {Pini}, \citenamefont
  {Galli},\ and\ \citenamefont {Bertaina}}]{bertaina_zenodo_TG_2021}%
  \BibitemOpen
  \bibfield  {author} {\bibinfo {author} {\bibfnamefont {M.}~\bibnamefont
  {Teruzzi}}, \bibinfo {author} {\bibfnamefont {C.}~\bibnamefont {Apostoli}},
  \bibinfo {author} {\bibfnamefont {D.}~\bibnamefont {Pini}}, \bibinfo {author}
  {\bibfnamefont {D.~E.}\ \bibnamefont {Galli}},\ and\ \bibinfo {author}
  {\bibfnamefont {G.}~\bibnamefont {Bertaina}},\ }\bibfield  {title} {\bibinfo
  {title} {{Data for: Evolution of static and dynamical density correlations in
  an one-dimensional soft-core gas from the Tonks-Girardeau limit to a
  clustering fluid}},\ }\href {https://doi.org/10.5281/zenodo.XXXXXXX}
  {10.5281/zenodo.XXXXXXX} (\bibinfo {year} {2021})\BibitemShut {NoStop}%
\end{thebibliography}
%
\end{document}